\documentclass[floats,aip,cha,reprint,superscriptaddress]{revtex4-1}

\usepackage{graphicx}
\usepackage{amsmath}   
\usepackage{amsfonts}
\usepackage{color}
\usepackage[margin=8mm,textfont=small,
position=top,
labelformat=empty,
caption=false,
justification=raggedright, singlelinecheck=false]{subfig}

\usepackage{placeins}

\usepackage{prettyref}
\newrefformat{sec}{Sec.~\ref{#1}}
\newrefformat{app}{Appendix~\ref{#1}}
\newrefformat{fig}{Fig.~\ref{#1}}
\newrefformat{tab}{Tab.~\ref{#1}}
\newrefformat{eq}{Eq.~(\ref{#1})}

\providecommand*{\fp}{\ensuremath{\mathrm{fp}}}
\providecommand*{\pointPSname}{\ensuremath{u}}
\providecommand*{\pointPS}{\ensuremath{\vec{\pointPSname}}}
\providecommand*{\fixedpoint}{\ensuremath{{\pointPS_{\fp}}}}

\providecommand*{\ui}{\text{i}}

\providecommand*{\valfixpt}[1]{\ensuremath{{\nu_{#1}^\fp}}}
\providecommand*{\nua}{\ensuremath{\valfixpt{1}}}
\providecommand*{\nub}{\ensuremath{\valfixpt{2}}}

\providecommand*{\zeroD}{\textsc{0d}}
\providecommand*{\oneD}{\textsc{1d}}
\providecommand*{\twoD}{\textsc{2d}}
\providecommand*{\threeD}{\textsc{3d}}
\providecommand*{\fourD}{\textsc{4d}}

\providecommand*{\CU}{\textsc{cu}}
\providecommand*{\EE}{\textsc{ee}}
\providecommand*{\EH}{\textsc{eh}}
\providecommand*{\HH}{\textsc{hh}}

\providecommand*{\Mone}{\ensuremath{{\cal M}_1^{\fp}}}
\providecommand*{\Mtwo}{\ensuremath{{\cal M}_2^{\fp}}}

\providecommand*{\pss}{\threeD{} phase-space slice}
\providecommand*{\psss}{\pss s}

\providecommand*{\onetori}{\oneD-tori}
\providecommand*{\onetorus}{\oneD-torus}
\providecommand*{\twotori}{\twoD-tori}
\providecommand*{\twotorus}{\twoD-torus}

\newcommand{\mT}{\ensuremath{m_{\text{N}}}}
\newcommand{\mL}{\ensuremath{m_{\text{L}}}}
\newcommand{\nuT}{\ensuremath{\nu_{\text{N}}}}
\newcommand{\nuL}{\ensuremath{\nu_{\text{L}}}}
\newcommand{\lamN}{\ensuremath{\lambda_{\text{N}}}}

\definecolor{blue}{rgb}{0.,0.5,1.}
\providecommand*{\bboxed}[1]{\fbox{\textcolor{blue}{#1}}}

\usepackage{hyperref}
\newcommand{\movierefall}{For a rotating view see
\href{http://www.comp-phys.tu-dresden.de/supp/}{
http://www.comp-phys.tu-dresden.de/supp/}.}
\newcommand{\movieref}[1]{For a rotating view of (#1) see
\href{http://www.comp-phys.tu-dresden.de/supp/}{
http://www.comp-phys.tu-dresden.de/supp/}.}

\makeatletter
\let\Hy@backout\@gobble
\makeatother

\begin{document}

\title{Bifurcations of families of 1D-tori in 4D symplectic maps}

\author{Franziska Onken}
\affiliation{Technische Universit\"{a}t Dresden, Institut f\"{u}r
  Theoretische Physik and Center for Dynamics, 01062 Dresden, Germany}
\affiliation{Max-Planck-Institut f\"{u}r Physik komplexer Systeme,
  N\"{o}thnitzer Stra{\ss}e 38, 01187 Dresden, Germany}

\author{Steffen Lange}
\affiliation{Technische Universit\"{a}t Dresden, Institut f\"{u}r
  Theoretische Physik and Center for Dynamics, 01062 Dresden, Germany}
\affiliation{Max-Planck-Institut f\"{u}r Physik komplexer Systeme,
  N\"{o}thnitzer Stra{\ss}e 38, 01187 Dresden, Germany}

\author{Roland Ketzmerick}
\affiliation{Technische Universit\"{a}t Dresden, Institut f\"{u}r
  Theoretische Physik and Center for Dynamics, 01062 Dresden, Germany}
\affiliation{Max-Planck-Institut f\"{u}r Physik komplexer Systeme,
  N\"{o}thnitzer Stra{\ss}e 38, 01187 Dresden, Germany}

\author{Arnd B\"acker}
\affiliation{Technische Universit\"{a}t Dresden, Institut f\"{u}r
  Theoretische Physik and Center for Dynamics, 01062 Dresden, Germany}
\affiliation{Max-Planck-Institut f\"{u}r Physik komplexer Systeme,
  N\"{o}thnitzer Stra{\ss}e 38, 01187 Dresden, Germany}

\date{\today}

\begin{abstract}
The regular structures of a generic \fourD{} symplectic map with a mixed phase
space are organized by one-parameter families of elliptic \onetori{}.
Such families show prominent bends, gaps, and new branches.
We explain these features in terms of bifurcations of the families when crossing
a resonance.
For these bifurcations no external parameter has to be varied. Instead,
the longitudinal frequency, which varies along the family, plays the role of
the bifurcation parameter.
As an example we study two coupled standard maps by visualizing the
elliptic and hyperbolic \onetori{} in a \pss{}, local \twoD{} projections, and
frequency space.
The observed bifurcations are consistent
with analytical predictions previously obtained for quasi-periodically forced
oscillators. Moreover, the new families emerging from such a bifurcation form
the skeleton of the corresponding resonance channel.
\end{abstract}

\maketitle

\noindent


\begin{quotation}
Bifurcations of invariant objects in dynamical systems
lead to an abrupt qualitative change of the dynamics
when an external parameter is varied smoothly.
In higher-dimensional systems regular tori in phase space are organized
around lower-dimensional tori.
Thus, the bifurcations of these lower-dimensional tori have strong implications
for the structures in phase space and the dynamics.
In this work the case of 4D maps is considered
for which families of \mbox{1D-tori} organize the regular 2D-tori, see
\prettyref{fig:phase_space}.
It turns out, that these families
undergo bifurcations without parameter variation. Hence, all stages of
a bifurcation can be seen at once,
which is visualized using 3D phase-space slices.
At the same time this leads to a better understanding of the phase-space
structure of resonance channels, which are of particular relevance for chaotic
transport in higher-dimensional systems.
\end{quotation}

\section{Introduction \label{sec:intro}}

The dynamics of higher-dimensional Hamiltonian systems is of interest to many
fields from physics, mathematics, and chemistry.
Such systems occur on scales ranging from atoms and
molecules~\cite{GekMaiBarUze2006, SchBuc1999, Kes2007, PasChaUze2008},
over chemical reactions~\cite{UzeJafPalYanWig2002,
  TodKomKonBerRic2005, ShoLiKomTod2007b, WaaSchWig2008, ManKes2014}
and particle accelerators~\cite{BazSibTur1997, VraIslBou1997, Pap2014}
to the solar system and galaxies~\cite{UdrPfe1988, Las1990, Cin2002,
PaeEft2015, JunZot2015b}.
Moreover, they exhibit new transport phenomena like Arnold
diffusion~\cite{Arn1964, KanBag1985, Loc1999, Cin2002, MalChi2010}.
Many of these examples are time-continuous systems, which can be reduced to
discrete-time maps: For example, an
autonomous Hamiltonian system with $f$ degrees
of freedom leads to a $2f$-dimensional continuous system that can be reduced
using energy conservation and a Poincar\'e section to a $(2f - 2)$-dimensional
symplectic map.
Moreover, in some cases the dynamics is directly described by such
maps~\cite{GasRic1989, GilEzr1991, DumLas1993}.
Thus, studying generic symplectic maps, which are numerically much more
convenient, helps to understand the dynamics of a wide range of systems.

\begin{figure}[b]
  \includegraphics{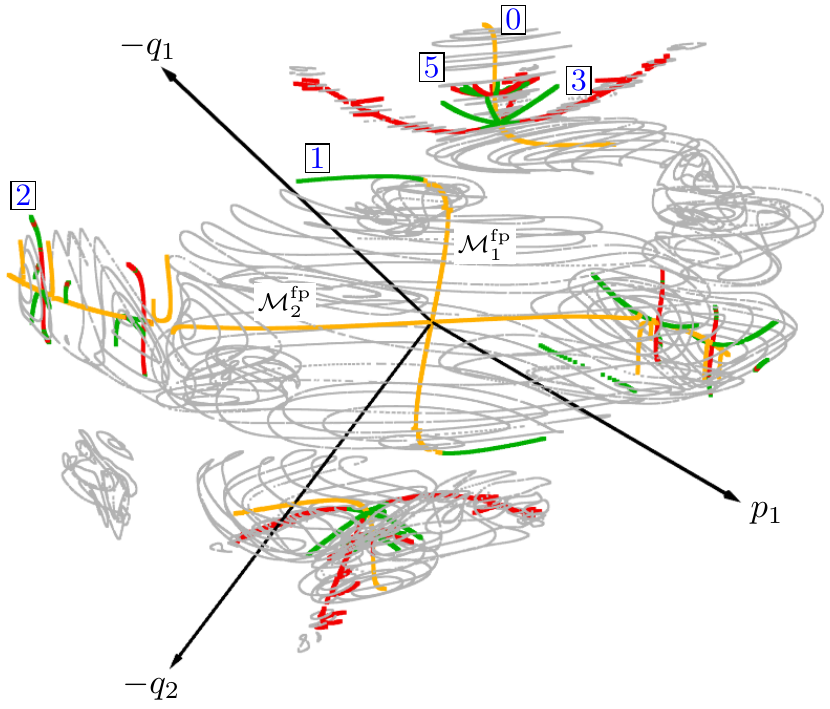}
  \caption{The \fourD{} phase space of two coupled standard maps
  \prettyref{eq:map},
  displayed in a \pss{}. Regular \twotori{} are shown as gray rings. The
  colored points forming lines are families of elliptic (orange, red) and
  hyperbolic (green) \onetori{}. The families \Mone{} and \Mtwo{} emanate from
  the \EE{} fixed point $\fixedpoint$ at the center $(p_1, q_1, q_2) = (0, 0.5, 0.5)$.
  Specific bifurcations are marked by blue boxed numbers and are
  discussed in detail in \prettyref{sec:bifs}. \movierefall{}}
  \label{fig:phase_space}
\end{figure}

Symplectic maps of dimension two are well understood, and generically possess
a mixed phase space in which regular and chaotic motion coexist.
The dynamics in  the \twoD{} phase space is organized by elliptic and
hyperbolic periodic orbits. The local behavior around these points is described
by the eigenvalues of the linearized dynamics. They contain either
the frequency $\nu$ describing the regular motion on \onetori{} around an
elliptic periodic orbit or the Lyapunov exponents of the stable and unstable
directions of a hyperbolic periodic orbit. If a parameter of the map is varied,
the frequency of an elliptic fixed point usually
changes and crosses rational values, $\nu = n/m$.
Depending on $m$ different types of bifurcations occur, namely
period-doubling ($|m| = 2$), touch-and-go ($|m| = 3, 4$),
and $m$-tupling ($|m| \ge 4$)
bifurcations~\cite{Mey1970, Bru1970, Alm1988, MaoDel1992, LebMou1999}.

The dynamics in maps of dimension four and higher has completely new
features as invariant regular tori have an insufficient dimension
to be barriers for the chaotic transport in phase space. This leads to
new effects like Arnold diffusion~\cite{Arn1964, KanBag1985,
Loc1999, Cin2002, MalChi2010} along so-called resonance channels.
In addition, the organization of the phase space of higher-dimensional maps is
more complicated~\cite{LanRicOnkBaeKet2014}: In general,
in a $2f$-dimensional map
elliptic $n$-dimensional tori~\cite{Eli1988, Poe1989} are
surrounded by (Cantor) families of elliptic $(n+1)$-dimensional
tori~\cite{JorVil1997, JorVil2001} with ${n = 0, 1, \dots, f - 1}$.
Lower-dimensional tori can additionally originate
from bifurcations of other lower-dimensional tori due to
resonances~\cite{GemTod1997, BroHanJorVilWag2003} or the break-up
of resonant higher-dimensional tori~\cite{Tod1994}.
Moreover, such bifurcations give rise to families of hyperbolic
lower-dimensional tori, also called whiskered tori~\cite{Arn1964, Zeh1976},
which are part of the so-called normally hyperbolic manifolds
(NHIMs). They play an important role for chaotic
transport~\cite{Wig1990, WaaSchWig2008}.
Note that the families of lower-dimensional tori are rather Cantor families as
they are interrupted by infinitely many gaps on finer and finer scales due to
resonances and their corresponding bifurcations.

We focus on \fourD{} symplectic maps in the generic regime far from
integrability which occur in many practical
applications~\cite{GilEzr1991, GekMaiBarUze2006, PasChaUze2008, LuqVil2011}. In
contrast, many former studies cover near-integrable systems, in which normal
form tools~\cite{Poe1989, Tod1994, BroHanJorVilWag2003} can be used.
In order to visualize such a generic \fourD{} phase space, we employ
\psss{}~\cite{RicLanBaeKet2014}, i.e.\ for every point of an orbit the
three remaining coordinates inside the slice are considered in a \threeD{} plot,
see \prettyref{fig:phase_space} for an example and
\prettyref{sec:basics} for a
brief introduction.
In a \fourD{} map the regular motion takes place on \twotori{}, some
of which are shown as gray
rings in \prettyref{fig:phase_space}. In addition, there are
\onetori{}, also
called invariant circles or fixed lines, which can be either
elliptic or hyperbolic~\cite{Gra1974, Zeh1976, JorVil1997}.
In the \pss{} they correspond to single points, which occur in
one-parameter families and thus lead to continuous looking lines.
Such families of elliptic \onetori{} are either attached to elliptic-elliptic
fixed points or arise from bifurcations, see orange and red lines in
\prettyref{fig:phase_space}, respectively. These families are
the skeleton of
the regular motion as each \onetorus{} is surrounded
by \twotori{} and
the families of \onetori{} form a hierarchy~\cite{LanRicOnkBaeKet2014}
analogously to the island-around-island hierarchy in \twoD{}
maps.
The hyperbolic \onetori{} occur in families as well, see green lines in
\prettyref{fig:phase_space}.
Furthermore, there are \zeroD{} periodic orbits, which can have stabilities
elliptic-elliptic (\EE{}), elliptic-hyperbolic
(\EH{}), hyperbolic-hyperbolic (\HH{}), and complex-unstable (\CU{}). They can
also undergo bifurcations under parameter variation~\cite{Pfe1985a,HowMac1987,
ConGio1988, KooMei1989a}, which, however, is not the subject of this study.

The aim of this paper is to explain the characteristics of the families of
elliptic \onetori{} in detail, namely their gaps, bends, and branches as
visible in \prettyref{fig:phase_space}.
It turns out that these structures can be understood as bifurcations of the
families caused by crossing resonances.
While bifurcations usually arise under parameter variation, we observe
all stages of a bifurcation in a single  phase space without varying any
parameter.
This is possible because the longitudinal, or intrinsic, frequency
of the \onetori{} varies smoothly along the family and plays the role of the
bifurcation parameter.
We observe that local \twoD{} projections of the
bifurcations in \psss{} remarkably resemble phase-space plots of bifurcations
of periodic orbits in \twoD{} maps.
This is consistent with normal form results for quasi-periodically forced
oscillators~\cite{BroHanJorVilWag2003} and investigations in \threeD{}
volume-preserving maps~\cite{DulMei2009}.
Moreover, the families of \onetori{} arising from a bifurcation due to a
crossing resonance are also the skeleton of this resonance channel.

This paper is organized as follows: In \prettyref{sec:basics} we introduce as
generic example system two coupled standard maps and give a short review of its
dynamics. In \prettyref{sec:bifs} bifurcations of families of \onetori{}
are categorized and illustrated in detail. This
local behavior is complemented by an investigation of global properties in
\prettyref{sec:global}. Finally, \prettyref{sec:summary} gives a summary and an
outlook. The special case of symmetry breaking bifurcations is described in
\prettyref{app:symm_breaking} and a short review of the used algorithm to
compute elliptic and hyperbolic \onetori{} is given in \prettyref{app:algo}.

\section{Dynamics in 4D maps \label{sec:basics}}

We consider as a generic example two coupled standard maps~\cite{Fro1972},
$(p_1, p_2, q_1, q_2) \mapsto (p_1', p_2', q_1', q_2')$,
    \begin{equation}
      \begin{aligned}
          p_1' &= p_1 + \frac{K_1}{2\pi} \sin(2\pi q_1')
                    + \frac{\xi_{12}}{2\pi} \sin(2\pi(q_1' + q_2')) \\
          p_2' &= p_2 + \frac{K_2}{2\pi} \sin(2\pi q_2')
                     + \frac{\xi_{12}}{2\pi} \sin(2\pi(q_1' + q_2')) \\
          q_1' &= q_1 + p_1 \\
          q_2' &= q_2 + p_2,
      \end{aligned}
      \label{eq:map}
    \end{equation}
where $p_{1, 2} \in [-0.5, 0.5)$ and $q_{1, 2} \in [0, 1)$ and periodic
boundary conditions are imposed in each coordinate. The resulting map is
symplectic. The parameters $K_1$ and $K_2$ control the nonlinearity of
the individual \twoD{} standard maps in $(p_1, q_1)$ and $(p_2, q_2)$,
respectively. The parameter $\xi_{12}$ introduces a coupling between the two
degrees of freedom.
We choose $K_1 = 2.25$, $K_2 = 3.0$ and $\xi_{12} = 1.0$, such that the system
is strongly coupled and far from integrability~\cite{RicLanBaeKet2014}. The
fixed point $\fixedpoint = (p_1, p_2, q_1, q_2) = (0, 0, 0.5, 0.5)$ is
of type \EE{}. The eigenvalues $(\lambda_1^\fp, \bar{\lambda}_1^\fp,
\lambda_2^\fp, \bar{\lambda}_2^\fp)$ of the linearized dynamics around
$\fixedpoint$ are $(\exp{(\pm \ui \ 2\pi \nua)}, \exp{( \pm \ui \ 2\pi \nub)})$
with $(\nua, \nub) = (0.30632, 0.12173)$. The mapping \prettyref{eq:map} has
been the subject of several studies~\cite{Pfe1985a, KanBag1985,
KooMei1989a, Las1993, JorOll2004}.

\begin{figure}[b]
  \includegraphics{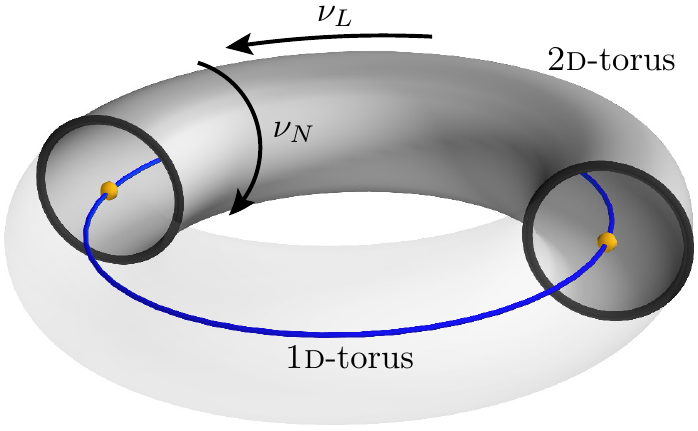}
  \caption{Sketch of structures in the \pss{}. A \twotorus{} (gray) appears
  generically as two rings (black) while a \onetorus{} (blue line) leads to two
  points (orange spheres). The longitudinal and the normal direction with
  corresponding frequencies \nuL{} and \nuT{} are indicated by arrows.}
  \label{fig:torus}
\end{figure}

An orbit started at some initial point in the \fourD{} phase space leads to a
sequence of points $(p_1, p_2, q_1, q_2)$ under the map \prettyref{eq:map}.
To visualize such an orbit we use a \pss{}~\cite{RicLanBaeKet2014}, defined
by thickening a \threeD{} hyperplane $\Gamma$ in the \fourD{} phase
space. Explicitly, we consider the slice defined by
\begin{align}
  \label{eq:slice-condition-in-coordinate}
  \Gamma_\varepsilon = \left\lbrace (p_1, p_2, q_1, q_2) \; \left|
      \rule{0pt}{2.4ex} \; |p_2 - p_2^*| \le \varepsilon \right. \right\rbrace
\end{align}
with $p_2^* = 0$ and $\varepsilon = 10^{-4}$ as it provides a good view of most
structures of the map \prettyref{eq:map}. Whenever a point of an orbit lies
within $\Gamma_\varepsilon$, the remaining coordinates $(p_1, q_1, q_2)$ are
displayed in a \threeD{} plot, see \prettyref{fig:phase_space}. Objects of
the \fourD{} phase space typically appear in the \pss{} with a dimension
reduced by one like sketched in \prettyref{fig:torus}, provided the object
intersects with the slice. Thus, a typical
\twotorus{} (gray) leads to two or more \oneD{} lines (black). A typical
\onetorus{} (blue line) leads to two or more points (orange spheres).
As \onetori{} occur in families, they lead to lines in the \pss{}, e.g.\ the
orange families in \prettyref{fig:phase_space}. A periodic
orbit will in
general not be visible, unless at least one of its points lies in the \pss{}.

\begin{figure}[t]
  \includegraphics{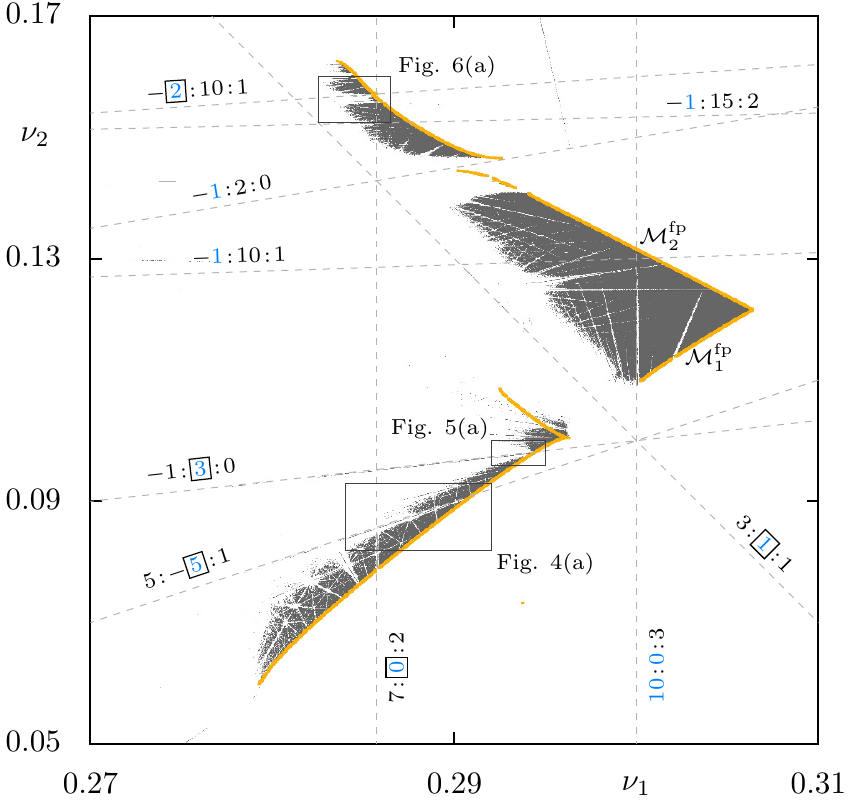}
  \caption{Frequency space of regular \twotori{} (gray points) of the coupled
standard maps, \prettyref{eq:map}. The rightmost tip $(\nua, \nub) = (0.30632,
0.12173)$ corresponds to the \EE{}
fixed point $\fixedpoint$. From there the two main families of elliptic
\onetori{} \Mone{} and \Mtwo{} (orange) emanate. Some resonances are shown as
dashed lines with labels $m_1:m_2:n$. In every label the factor $|\mT|$
is marked in blue. The boxed
values correspond to the bifurcations labeled in
\prettyref{fig:phase_space}.
The three black boxes indicate magnifications shown in
\prettyref{fig:res_5_-5_1_fs}, \prettyref{fig:res_-1_3_0_fs} and
\prettyref{fig:res_-2_10_1_fs}.
}
  \label{fig:freq_space}
\end{figure}

In order to relate bifurcations to resonances, phase space objects are
represented in frequency space. Using frequency analysis~\cite{Las1990,
BarBazGioScaTod1996} we associate with each \twotorus{} its two fundamental
frequencies $(\nu_1, \nu_2) \in [0, 1)^2$, see \prettyref{fig:torus}.
These are displayed in the frequency space, see
\prettyref{fig:freq_space}, where the gray points represent \twotori{} obtained
by starting $10^8$ initial conditions with randomly chosen $p_1, p_2
\in [-0.2, 0.2]$, $q_1, q_2 \in [0.3, 0.7]$ in the \fourD{} phase space. Each
frequency pair is calculated from $N=4096$ iterations using a Fourier method
(Sec. 4.2.4 in Ref.~\cite{BarBazGioScaTod1996}). Note that there are no
rotational tori in the system at the chosen parameters.
To decide whether an orbit is regular we use the
frequency criterion $\max{\left(|\nu_1 - \tilde{\nu}_1|,|\nu_2 -
\tilde{\nu}_2|\right)} < 10^{-7}$,
where the second frequency pair $(\tilde{\nu}_1, \tilde{\nu}_2)$ is calculated
from
$N$ further iterations. This leads to nearly $3 \cdot 10^6$ regular \twotori{}.
The frequencies $(\nu_1, \nu_2)$ are only defined up to a unimodular
transformation~\cite{Bor1927, GekMaiBarUze2007}.
Starting from the \EE{} fixed point with frequencies $(\nua, \nub)$ we have used
the \pss{} to obtain a consistent representation
in frequency space by applying appropriate unimodular transformations
such that neighboring regions in phase space
are mapped to neighboring regions in frequency space~\cite{RicLanBaeKet2014}.

The frequency space is covered by resonance lines $m_1:m_2:n$, on
which the frequencies fulfill
\begin{align}
  \label{eq:resonance-driven}
  m_1 \cdot \nu_1 + m_2 \cdot \nu_2 = n
\end{align}
where $m_1, m_2, n$ are integers without common divisor. Some resonance lines
that are relevant in the following are
displayed in \prettyref{fig:freq_space} as dashed lines.

On a \onetorus{} the dynamics is described by only one longitudinal frequency
\nuL{} which is also called intrinsic or internal frequency or rotation number.
If the \onetorus{}
is elliptic, it is surrounded by \twotori{}. For a sequence of such \twotori{}
converging to a \onetorus{}, one of the two frequencies $(\nu_1, \nu_2)$
converges towards the longitudinal frequency \nuL{}, see
\prettyref{fig:torus}. The limit of the other frequency corresponds to the
normal frequency \nuT{} describing the rotation of points normal to the
elliptic \onetorus{}~\cite{JorVil1997}. Therefore, the families of \onetori{}
are represented by the edges in frequency space, see \prettyref{fig:freq_space}.
The normal frequency \nuT{} may also be obtained from the linearized motion
around the \onetorus{}~\cite{Jor2001}.

There are two families \Mone{}, \Mtwo{} of elliptic \onetori{} emanating from
the \EE{} fixed point $\fixedpoint$, shown as orange lines in
\prettyref{fig:freq_space}.
On the family \Mone{} the frequency $\nu_1$ corresponds to the longitudinal
frequency $\nuL$ and $\nu_2$ to the normal one $\nuT$, i.e.\ a resonance $m_1 :
m_2 : n$ crossing \Mone{} corresponds to $\mL{} : \mT{} : n$.
On \Mtwo{} it is the other way around, i.e.\  $m_1 : m_2 : n$ corresponds to
$\mT{} : \mL{} : n$. As discussed in the following section, \mT{} is the
decisive parameter for the categorization of bifurcations of families of
\onetori{}.
The respective values of $\mT{}$ are marked as blue numbers
in \prettyref{fig:freq_space}.

\section{\label{sec:bifs}Bifurcations of families of 1D-tori}

In this section we explain the structures that occur along the main families
\Mone{} and \Mtwo{} in \prettyref{fig:phase_space}, i.e.\ new
emerging
branches, gaps, and strong bends, by bifurcations. These bifurcations occur as
the frequencies \nuL{} and
\nuT{}, which vary smoothly along every family, cross resonance
lines, see \prettyref{fig:freq_space}. To study the structures at a
bifurcation we compute the newly emerging elliptic and hyperbolic
families of \onetori{}, see \prettyref{app:algo} for details, and relate all
involved objects in phase space and in frequency space.

Bifurcations of \onetori{} were also investigated for quasi-periodically forced
oscillators~\cite{BroHanJorVilWag2003} and \threeD{} volume-preserving
maps~\cite{DulMei2009}, and it was found that the normal coefficient \mT{} of
the crossing resonance is the relevant parameter to categorize such
bifurcations.
In the following, we discuss the different cases of \mT{} to explain the
structures in phase space starting with the generic case $|\mT| \ge
4$ (branches) and then continue with $|\mT| = 3$ (branches), $|\mT| = 2$
(gaps), $|\mT| = 1$ (bends) and $|\mT| = 0$ (gaps). The resonances used as
examples are marked by dashed lines in \prettyref{fig:freq_space} with labels
in which the values of \mT{} are blue and boxed.
Further examples have been investigated~\cite{Onk2015}.
There also exist symmetry breaking bifurcations, which are discussed in
\prettyref{app:symm_breaking}.

This categorization of bifurcations is analogous to \twoD{} maps, where the
resonance condition reads $m \cdot \nu = n$ and $m$ is the relevant
number to categorize the
bifurcations~\cite{Mey1970, Bru1970, Alm1988, LebMou1999}.
It turns out that for $\mT{} = m$ the features of the bifurcations of families
of \onetori{} in \fourD{} maps resemble bifurcations of fixed points in \twoD{}
maps.

In the following we refer to the family of \onetori{} which is crossed
by a resonance as \emph{main family}.
Families of elliptic \onetori{} created in a bifurcation also undergo
bifurcations which leads to a whole hierarchy~\cite{LanRicOnkBaeKet2014}.
All bifurcations occur independent of the hierarchy and of the origin of the
family, i.e.\ whether they arise from fixed points or from other bifurcations.
Thus, without loss of generality we concentrate on the first level of the
hierarchy, i.e.\ in the following the main family is always either \Mone{} or
\Mtwo{} indicated in orange.

In the visualizations we observe that the gaps within the families of
\onetori{} created in a bifurcation get very small close to the main family.
It is known that for elliptic families of \onetori{} attached at \EE{}
fixed points these gaps get exponentially small with the distance to the fixed
point~\cite{JorVil1997, JorOll2004}. Thus, we conjecture that also
the gaps in the families of \onetori{} arising from a bifurcation get
exponentially small close to the main family.

To visualize and analyze the bifurcations, we use in addition to the frequency
space and the \pss{} also local \twoD{} projections of the \pss{}. They are
useful to relate the results to both \twoD{} bifurcations and normal form
results. The basic methods and concepts are introduced in detail in the
following section.

\subsection{Branches: $\boldmath{|\mT| \ge 4}$ \label{sec:branches}}

For $|\mT| \ge 4$ one obtains generic bifurcations of a family of \onetori{} in
which one elliptic and one hyperbolic family of \onetori{} are created.
As the geometry of the \onetori{} is dictated~\cite{Tod1994, DulMei2009,
LanRicOnkBaeKet2014} by the crossing resonance condition $\mL : \mT : n$,
each \onetorus{} consists of $\gcd(\mL, \mT)$ disjoint but dynamically
connected loops.
Furthermore, the families of \onetori{} appear in a typical \pss{} as
alternating $|\mT|$ elliptic and $|\mT|$ hyperbolic branches.

As an example we discuss in detail the $5:-5:1$ resonance intersecting \Mone{},
i.e.\ $|\mT|= 5$, see \prettyref{fig:freq_space}. This resonance
causes the red and green branches in phase space marked by \bboxed{5} in
\prettyref{fig:phase_space}.

\begin{figure*}[htbp]
  \subfloat[][]{
    \includegraphics{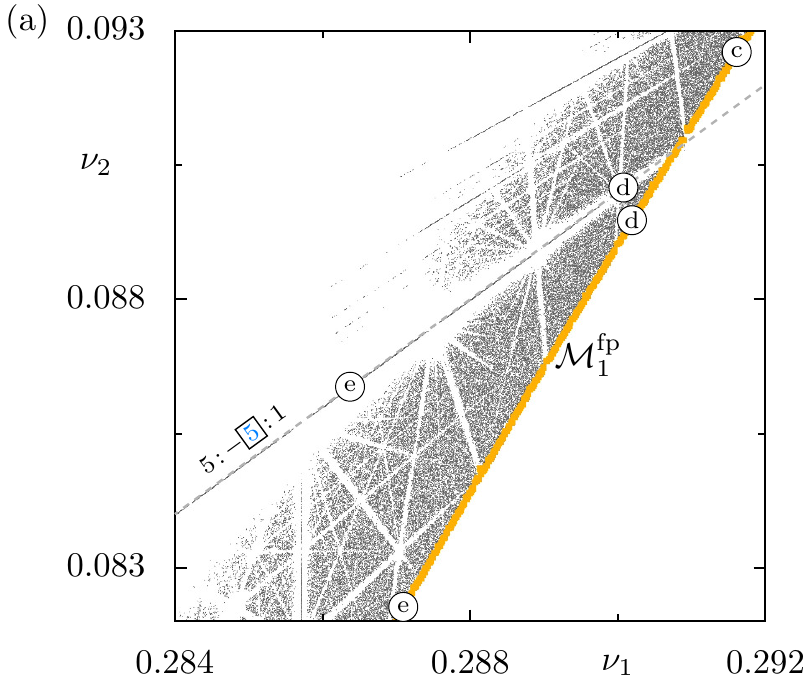}
    \label{fig:res_5_-5_1_fs}
  }
  \subfloat[][]{
    \includegraphics{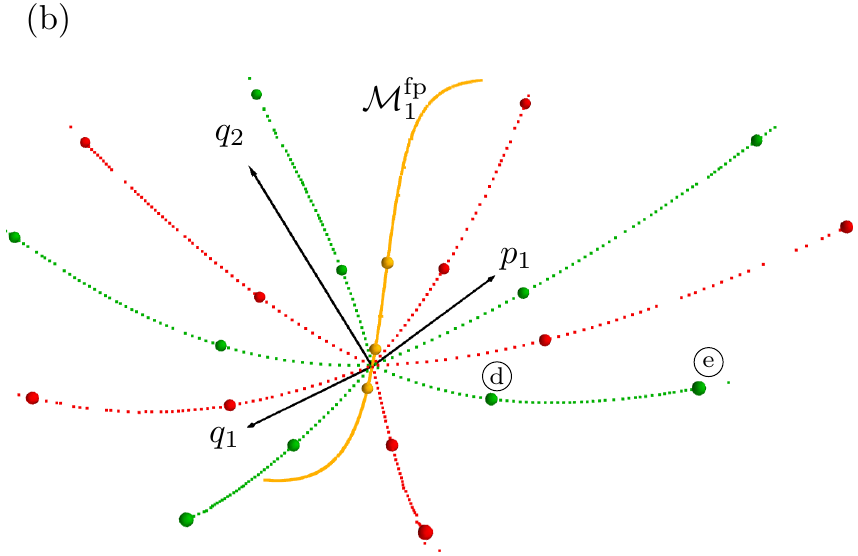}
    \label{fig:res_5_-5_1_ps}
  } \\
  \subfloat[][]{
    \includegraphics{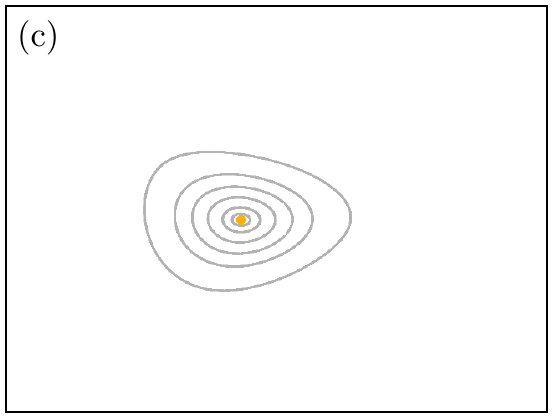}
    \label{fig:res_5_-5_1_2Dplane1}
  }
  \subfloat[][]{
    \includegraphics{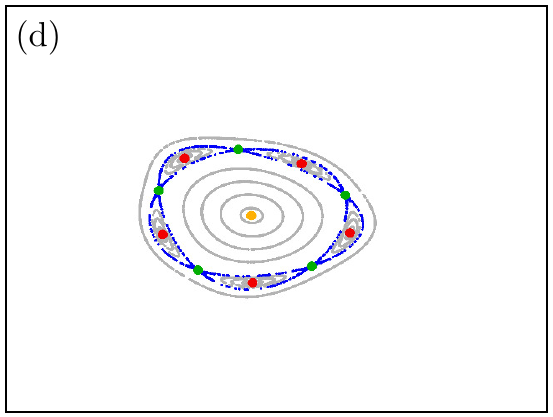}
    \label{fig:res_5_-5_1_2Dplane2}
  }
  \subfloat[][]{
    \includegraphics{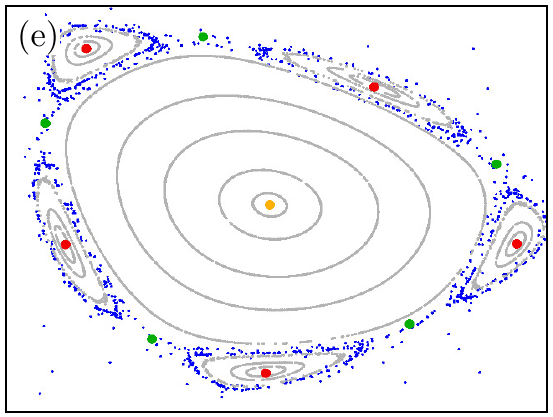}
    \label{fig:res_5_-5_1_2Dplane3}
  } \\
  \subfloat[][]{
    \includegraphics{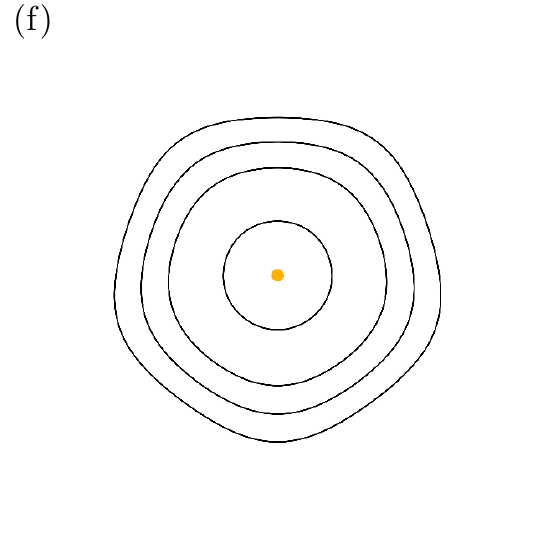}
    \label{fig:res_5_-5_1_normform1}
  }
  \subfloat[][]{
    \includegraphics{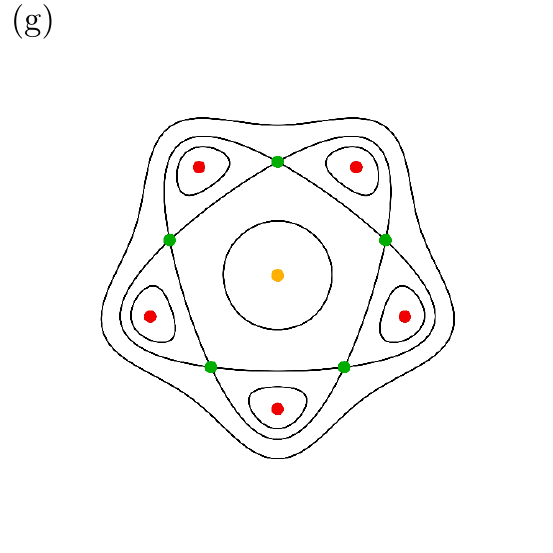}
    \label{fig:res_5_-5_1_normform2}
  }
  \subfloat[][]{
    \includegraphics{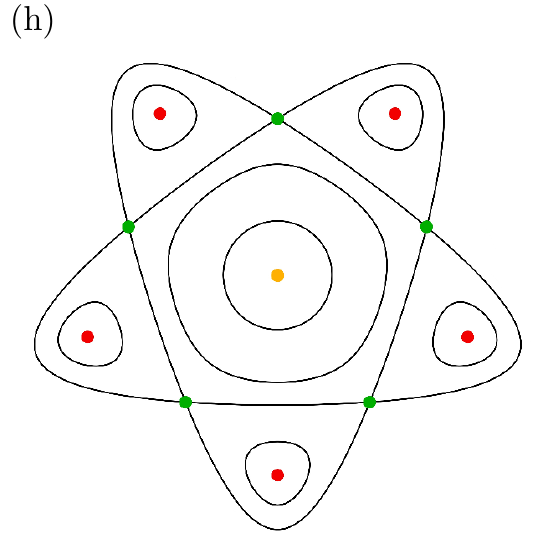}
    \label{fig:res_5_-5_1_normform3}
  }
  \caption{Visualization of the $5:-\bboxed{5}:1$ resonance intersecting
\Mone{}.
  (a) Zoom of frequency space,
\prettyref{fig:freq_space}, with resonance line
(dashed gray line), regular \twotori{} (gray) and a part of \Mone{}(orange).
  (b) Magnified and rotated view of the structure labeled with \bboxed{5} in the
\pss{} in \prettyref{fig:phase_space} with elliptic (red) and
hyperbolic \onetori{} (green). The orange line is a part of \Mone{}.
  (c)--(e) Sequence of local \twoD{} projections from the \pss{} onto a plane
with normal vector $\vec{n} =
(-0.338, -0.644, 0.687)$ illustrating the stages of the bifurcation with
regular \twotori{} (gray rings), elliptic \onetori{} (red, orange), and
hyperbolic \onetori{} (green). The \onetori{} appearing in the local \twoD{}
projections (colored points) are marked by encircled letters
\textcircled{\small c} to \textcircled{\small e} in (a) and by larger spheres in
(b).
  (f)--(h) Contour plots of the simplified normal
form~\cite{BroHanJorVilWag2003}
$f(I, \phi; \lambda, \mu) = \lambda I - I^2 + \mu^{m-3} A (2I)^{m/2}
\cos(m \phi)$ with parameters
 $m=5$, $\mu^{m-3} \cdot A \cdot 2^{m/2} = -0.4$
(f) before, $\lambda = -0.2$,
(g) right after, $\lambda = 0.2$, and (h) further away from the
bifurcation, $\lambda = 0.4$. The colored points mark the equilibria.
 \movieref{b}
}
  \label{fig:res_5_-5_1}
\end{figure*}

This example is visualized in several ways in
\prettyref{fig:res_5_-5_1}:
Figure \ref{fig:res_5_-5_1_fs} shows a
zoom of the frequency space of \prettyref{fig:freq_space} close to the crossing
point of the resonance and the main family~\Mone{}. In
\prettyref{fig:res_5_-5_1_ps} the structure
is shown in a \pss{}, which is a rotated magnification of
\prettyref{fig:phase_space}. There one
can see five hyperbolic (green) and five elliptic (red) branches emerging out
of the main family \Mone{} (orange).
Note that the branches consist of individual points, where each point
corresponds to an intersection of a \onetorus{} with the \pss{}.
Every \onetorus{} of these branches consists of $\gcd(5, -5) = 5$ disjoint
loops.

\begin{figure*}[p]
  \subfloat[][]{
    \includegraphics{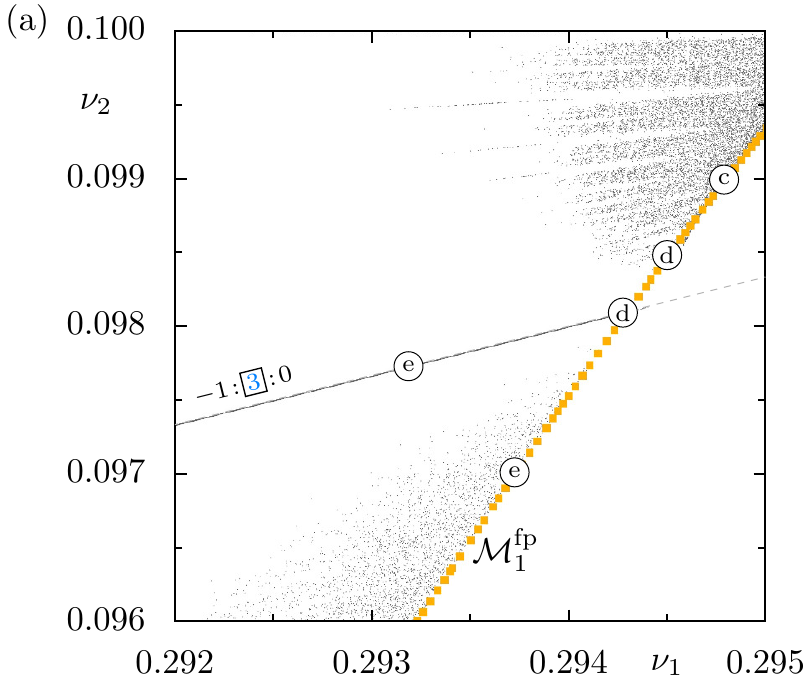}
    \label{fig:res_-1_3_0_fs}
  }
  \subfloat[][]{
    \includegraphics{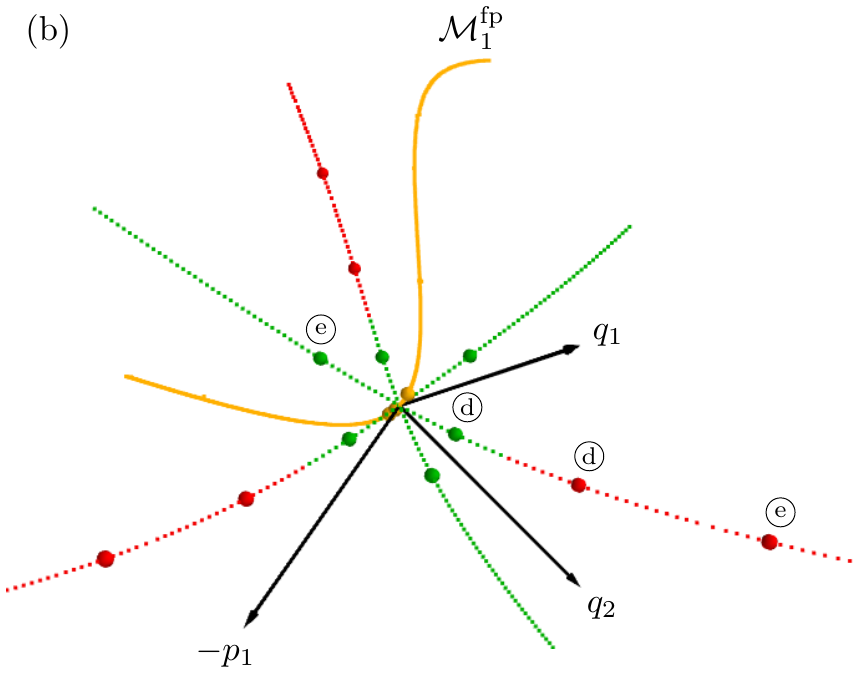}
    \label{fig:res_-1_3_0_ps}
  } \\
  \subfloat[][]{
    \includegraphics{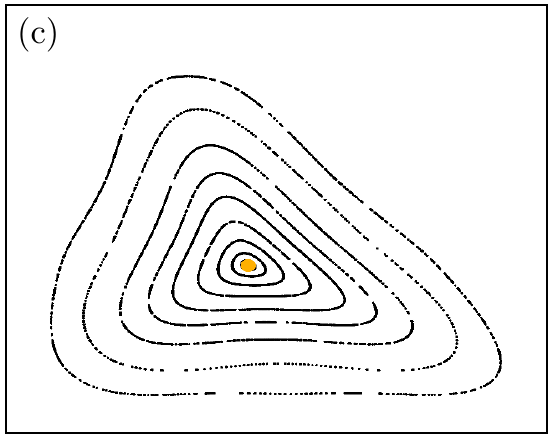}
    \label{fig:res_-1_3_0_2Dplane1}
  }
  \subfloat[][]{
    \includegraphics{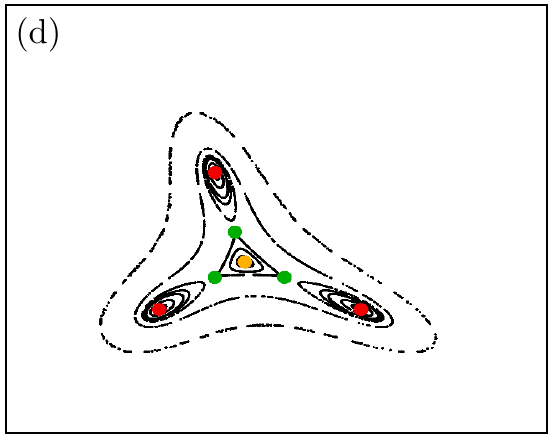}
    \label{fig:res_-1_3_0_2Dplane2}
  }
  \subfloat[][]{
    \includegraphics{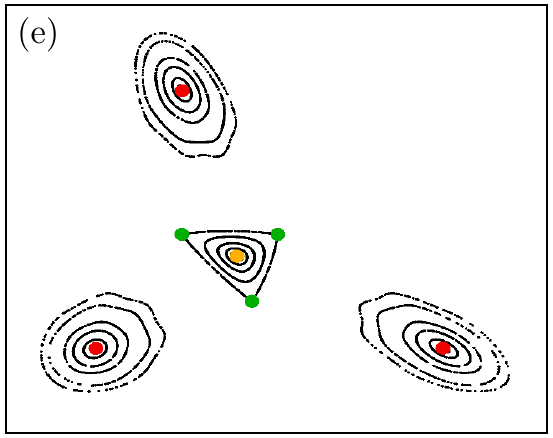}
    \label{fig:res_-1_3_0_2Dplane3}
  } \\
  \subfloat[][]{
    \includegraphics{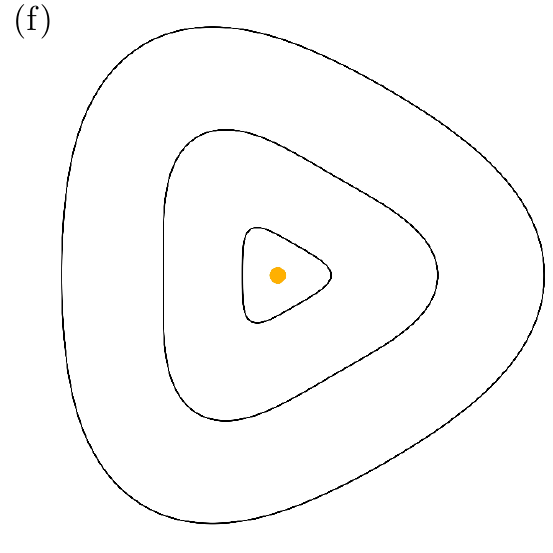}
    \label{fig:res_-1_3_0_normform1}
  }
  \subfloat[][]{
    \includegraphics{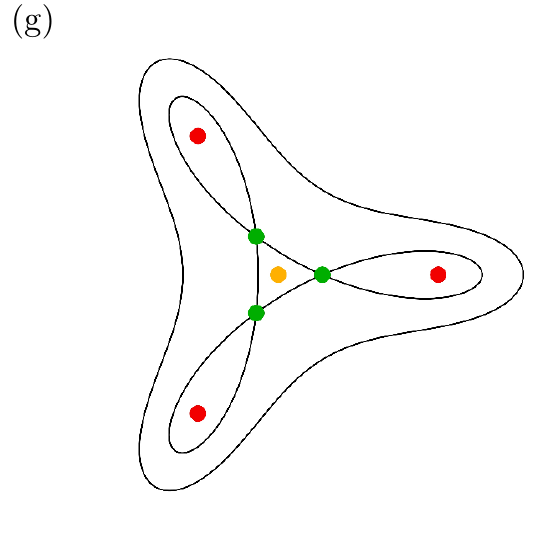}
    \label{fig:res_-1_3_0_normform2}
  }
  \subfloat[][]{
    \includegraphics{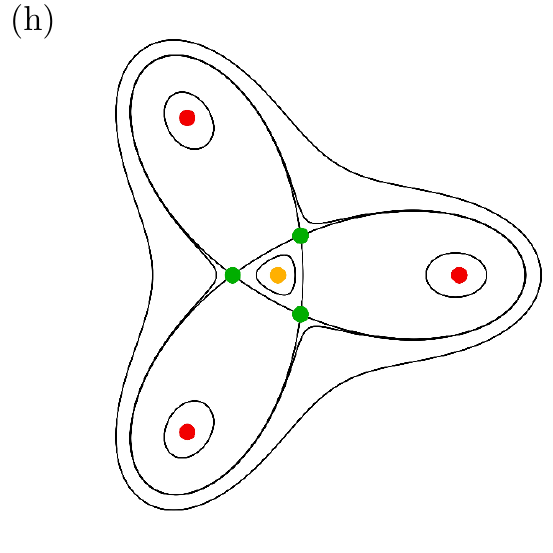}
    \label{fig:res_-1_3_0_normform3}
  }
  \caption{Visualization of the $-1:\bboxed{3}:0$ resonance similar to
\prettyref{fig:res_5_-5_1}.
  (c)--(e) Local \twoD{} projections with normal vector $\vec{n} =
(-0.336, -0.672, 0.66)$.
  (f)--(h) Contour plots of the normal form~\cite{BroHanJorVilWag2003}
  $f(p, q; \lambda, A) = \lambda \frac{p^2 + q^2}{2} - \left(\frac{p^2 +
q^2}{2} \right)^2 + A (q^3 - 3 p^2 q)$, $A = 0.05$, (f)
before, $\lambda=-0.016$,
(g) after the saddle-node part, $\lambda=-0.004$, and (h) after the
touch-and-go bifurcation, $\lambda=0.01$.
\movieref{b}
}
  \label{fig:res_-1_3_0}
\end{figure*}

For a clear visualization of the stages of the bifurcation, we consider
in Figs.~\ref{fig:res_5_-5_1}(c)--(e) \emph{local 2D projections} from the
\pss{} of \prettyref{fig:res_5_-5_1_ps} onto \twoD{} planes.
They show
several orbits started in the respective plane, such that
the \twotori{} are represented as gray rings and the \onetori{} as colored
points. Note that the chosen \twoD{} projections help to focus on a specific
part of the \pss{}. The normal vector of the planes for the shown \twoD{}
projections is chosen along the family \Mone{} of \onetori{} (orange).
The positions of the local \twoD{} projections are indicated in
\prettyref{fig:res_5_-5_1_fs} by the frequencies of the elliptic \onetori{}
involved in Figs.~\ref{fig:res_5_-5_1}(c)--(e), marked with encircled
characters \textcircled{\small c} to \textcircled{\small e}, and
in \prettyref{fig:res_5_-5_1_ps} by enlarged points.
By this variation of the position of the local \twoD{} projection along \Mone{}
the different stages of the bifurcation become visible without varying any
parameter of the map.

The resulting \twoD{} plots look remarkably similar to a period-5-tupling
bifurcation in \twoD{} maps.
However, the interpretation is different: an isolated point in the local
\twoD{} projections corresponds to a \onetorus{} while a closed line
corresponds to a
\twotorus{}. For a \twoD{} map these objects would correspond to a periodic
point and a regular torus, respectively.

These observations are consistent with normal form results obtained for
parametrized families of quasi-periodically forced Hamiltonian
oscillators~\cite{BroHanJorVilWag2003}. Figures~\ref{fig:res_5_-5_1}(f)--(h)
show contour plots of the corresponding normal form for different stages of the
bifurcation. The stable and unstable equilibrium points are marked in red and
green, respectively. Figures~\ref{fig:res_5_-5_1}(f)--(h)
demonstrate how an ``island chain'' consisting of five stable and five
unstable equilibria emerges out of the \onetorus{} in the center.
After the bifurcation the distance of the islands to the center increases.
The same behavior is observed in Figs.~\ref{fig:res_5_-5_1}(c)--(e) along the
main family.
This resemblance supports the conjecture~\cite{BroHanJorVilWag2003}
that the normal form results obtained for quasi-periodically forced oscillators
also apply to generic fully coupled systems. Note that the normal form does not
describe any gaps in the arising families, which are visible in
\prettyref{fig:res_5_-5_1_ps}.

It is worth emphasizing that the new elliptic and hyperbolic families of
\onetori{} shown in \prettyref{fig:res_5_-5_1_ps} and their
surrounding form the
resonance channel in phase space. In frequency space in
\prettyref{fig:res_5_-5_1_fs} this whole phase-space region collapses to the
resonance line $5:-5:1$.

In \prettyref{fig:res_5_-5_1_2Dplane2} and
\prettyref{fig:res_5_-5_1_2Dplane3} chaotic orbits with initial conditions
inside the resonance channel are included as blue points.
In \prettyref{fig:res_5_-5_1_2Dplane2} the chaotic orbits in the resonance
channel are effectively confined by KAM tori, such that they cannot
escape the channel on short time scales.
In contrast the chaotic orbits in \prettyref{fig:res_5_-5_1_2Dplane3}
clearly escape from the channel.
These two regimes can also be seen in frequency space,
\prettyref{fig:res_5_-5_1_fs}.
Near the bifurcation (upper \textcircled{\raisebox{-0.9pt} {d}}), i.e.\
close to the orange main family, the channel is embedded in gray \twotori{},
but further away (upper \textcircled{e}), such confining
\twotori{} cease to exist.

\subsection{Branches: $\boldmath{|\mT| = 3}$ \label{sec:mN3}}

Bifurcations with $|\mT| = 3$ are more complicated: Initially,
an elliptic and a hyperbolic family of \onetori{} emerge in a saddle-node
bifurcation.
The hyperbolic family collapses at the main family and reemerges from it.
Thus, this region is considerably deprived of regular tori.
The families appear in a \pss{} as three hyperbolic and three elliptic branches.

As an example we consider the $-1:3:0$ resonance intersecting \Mone{}. In
\prettyref{fig:res_-1_3_0} a series of plots analogous to
\prettyref{fig:res_5_-5_1} is shown. The $-1:3:0$ resonance leads to elliptic
(red) and hyperbolic (green) branches in phase space, marked by \bboxed{3} in
\prettyref{fig:phase_space}, and
a large open resonance channel in frequency space, see
\prettyref{fig:res_-1_3_0_fs}.
Especially, in the \pss{} in \prettyref{fig:res_-1_3_0_ps} one
recognizes that the elliptic branches (red) do not start at the main family
\Mone{} (orange) but at the saddle-node bifurcation, i.e.\ the point where the
hyperbolic (green) and the elliptic (red) part of the branch meet.

The stages of this bifurcation can again be understood by means of the
local \twoD{} projections. In Figs.~\ref{fig:res_-1_3_0}(c)--(e)
projections of orbits started in the respective \twoD{} plane are
shown. The positions of the local \twoD{} projections are indicated in
\prettyref{fig:res_-1_3_0_fs} by circled characters and in
\prettyref{fig:res_-1_3_0_ps} by larger spheres.
These pictures not only resemble all stages of a \twoD{} touch-and-go period
tripling bifurcation, but also match qualitatively with normal form
predictions~\cite{BroHanJorVilWag2003}, which are shown in
Figs.~\ref{fig:res_-1_3_0}(f)--(h).
Note that this normal form of the period-tripling already includes the
saddle-node bifurcation in contrast to the corresponding one
for \twoD{} area-preserving maps~\cite{Mey1970,Bru1970},
for which, despite of this, the period-tripling is still
commonly observed to be preceded by the saddle-node
bifurcation~\cite{MaoDel1992,Sch1997a,Bae1998}.

In both series (c)--(e) and (f)--(h) the first figure shows the phase space
before the bifurcation with just deformed \twotori{}. The second one shows the
stage after the saddle-node bifurcation including the three new hyperbolic
families of \onetori{} (green) and the three islands each centered around a
family of elliptic \onetori{} (red). The last stage is the ``touch-and-go''
part, where the three hyperbolic branches (green) meet at the main family
\Mone{} and cross each other. The third figures show the phase space
after this stage, i.e.\ where all branches just continue to spread out from the
main family \Mone{}.

\subsection{Gaps: $\boldmath{|\mT| = 2}$ \label{sec:mN2}}

Bifurcations with $|\mT| = 2$ cause gaps in the main family of elliptic
\onetori{} in the sense that the family contains a segment of hyperbolic
\onetori{}:
Firstly, the main family of elliptic \onetori{} changes its stability to
hyperbolic while one elliptic family emerges, appearing as two branches in the
\pss{}. Secondly, the main
family of now hyperbolic \onetori{} becomes elliptic again and a hyperbolic
family emerges, also appearing as two branches in the \pss{}.
This bifurcation is similar to a combination of a direct
and an inverse period-doubling bifurcation in \twoD{} maps.

\begin{figure}[htbp]
  \subfloat[][]{
    \includegraphics{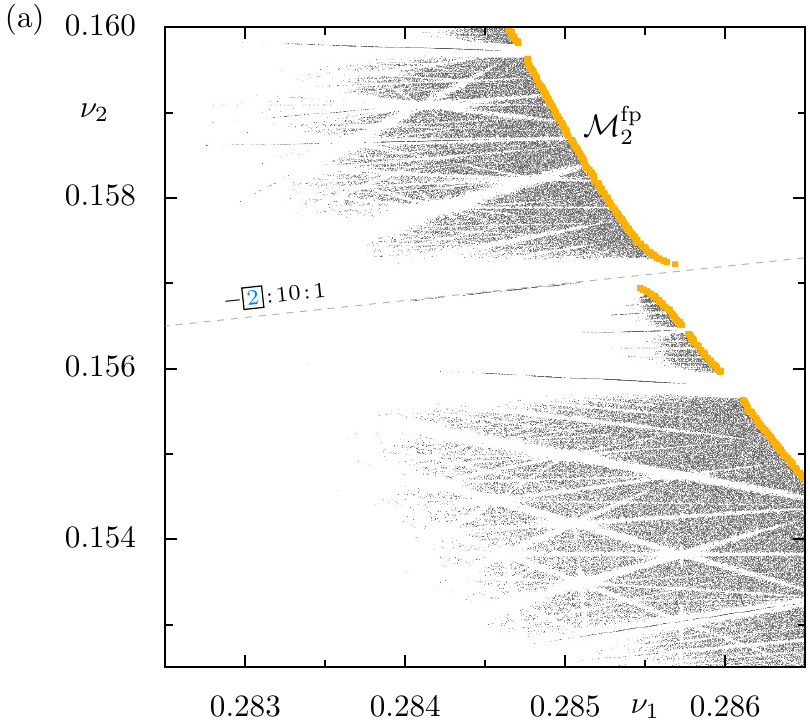}
    \label{fig:res_-2_10_1_fs}
  } \\
  \subfloat[][]{
    \includegraphics{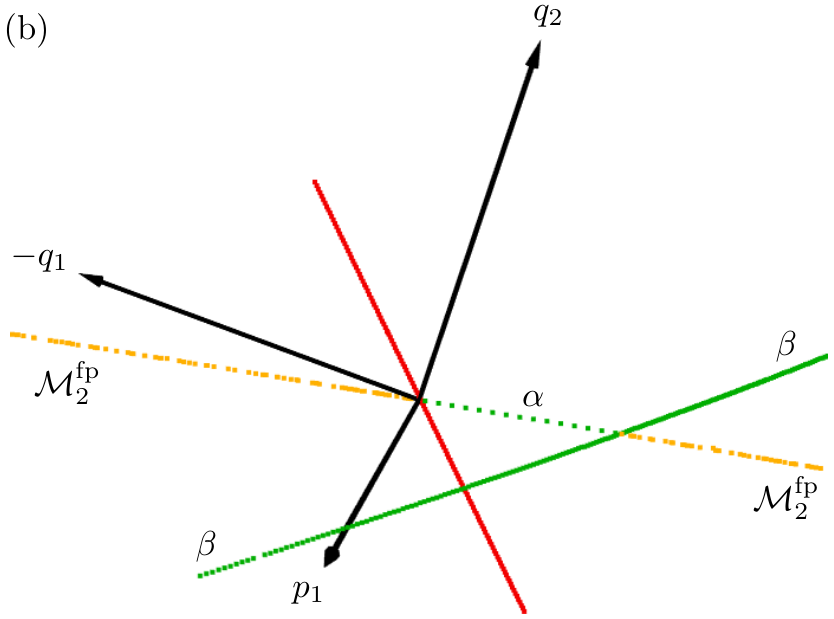}
    \label{fig:res_-2_10_1_ps}
  } \\
  \subfloat[][]{
    \includegraphics{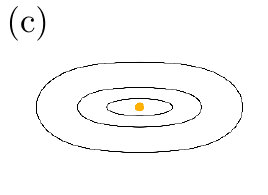}
    \label{fig:res_-2_10_1_normform1}
  }
  \subfloat[][]{
    \includegraphics{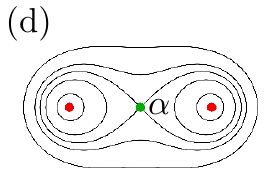}
    \label{fig:res_-2_10_1_normform2}
  }
  \subfloat[][]{
    \includegraphics{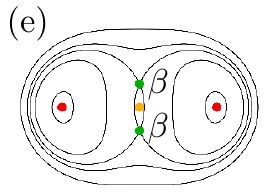}
    \label{fig:res_-2_10_1_normform3}
  }
  \caption{Visualization of the $-\bboxed{2}:10:1$ resonance as in
\prettyref{fig:res_5_-5_1}.
  (b) The orange family \Mtwo{} temporarily changes stability to hyperbolic
($\alpha$) due to the $|\mT| = 2$ bifurcation. At each end of the hyperbolic
part ($\alpha$) new elliptic (red) and hyperbolic (green, $\beta$) families
emerge.
(c)--(e) Contour plots of the normal form~\cite{BroHanJorVilWag2003} $f(p, q;
\delta, A) = (\delta + a) \frac{p^2 + q^2}{2} - \left(\frac{p^2 +
q^2}{2} \right)^2 + A (q^2 - p^2)$, $A = 0.05$, $a = 0.1$, (c) before,
$\delta=-0.21$,
(d) after the first period-doubling, $\delta=-0.1$, and (e) after the second
period-doubling bifurcation, $\delta=0.02$.
\movieref{b}
}
  \label{fig:res_-2_10_1}
\end{figure}

As an example we consider the $-2:10:1$ resonance, marked by
\bboxed{2} in Figs.~\ref{fig:phase_space} and
\ref{fig:freq_space},
which intersects \Mtwo{}.
Figure~\ref{fig:res_-2_10_1_fs} is a zoom of the frequency space in
\prettyref{fig:freq_space} and shows a bend of the main family (orange), which
appears to ``converge'' to the resonance line from either side.
Figure~\ref{fig:res_-2_10_1_ps} shows the structure
in the \pss{}. The main family \Mtwo{} (orange) is interrupted by a hyperbolic
part (green, marked by $\alpha$) and when turning hyperbolic a
family of elliptic \onetori{} (red) emerges. When the main family turns
elliptic again, a family of hyperbolic \onetori{} (green, marked by $\beta$)
arises.

For this bifurcation we have not been able to find a sequence of local \twoD{}
projections which clearly shows the different stages. Thus, we now use the
intuition gained from bifurcations with $|\mT| \ge 3$
(sections \ref{sec:branches} and \ref{sec:mN3}) to
understand the connection between the \pss{} in
\prettyref{fig:res_-2_10_1_ps}
and normal form plots~\cite{BroHanJorVilWag2003} in
Figs.~\ref{fig:res_-2_10_1}(c)--(e).
In these figures the stages of the bifurcation are shown from left to right:
(c) shows the phase space before the bifurcations, (d) after the first (direct)
period doubling and (e) after the second (inverse) period doubling bifurcation.
In (e) the emerging hyperbolic branches (green) lie on a line perpendicular to
the line of the new elliptic branches (red). This geometric
relation is also recognizable in the \pss{}, see rotating view of
\prettyref{fig:res_-2_10_1_ps}.

\subsection{Bends: $\boldmath{|\mT| = 1}$ \label{sec:mN1}}

\begin{figure}[b]
  \subfloat[][]{
    \includegraphics{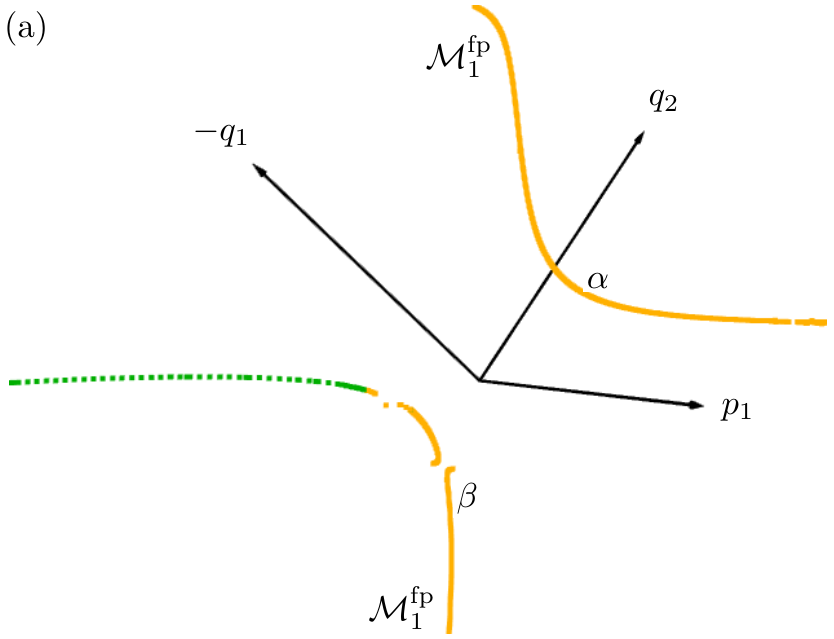}
    \label{fig:res_3_1_1_ps}
  } \\
  \subfloat[][]{
    \includegraphics{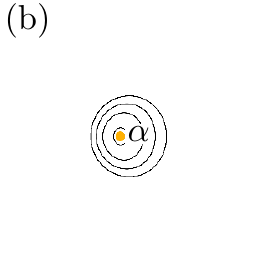}
    \label{fig:res_3_1_1_normform1}
  }
  \subfloat[][]{
    \includegraphics{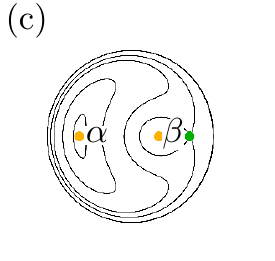}
    \label{fig:res_3_1_1_normform2}
  }
  \subfloat[][]{
    \includegraphics{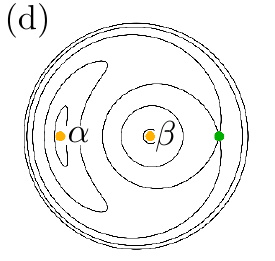}
    \label{fig:res_3_1_1_normform3}
  }
  \caption{Visualization of the $3:\bboxed{1}:1$ resonance as in
\prettyref{fig:res_5_-5_1}.
  (a) The orange family labeled with $\alpha$ is a part of
\Mone{} and bends away due to the $|\mT| = 1$ bifurcation.
  (b)--(d) Contour plots of the normal form~\cite{BroHanJorVilWag2003} $f(p, q;
\delta, A) = \delta \frac{p^2 + q^2}{2} - \left(\frac{p^2 +
q^2}{2} \right)^2 + A q $, $A = -0.006$, (b) before, $\delta=-0.04$,
(c) right after, $\delta=0.08$, and (d) later after, $\delta=0.16$, the
bifurcation. (d) shows how the elliptic
branch $\beta$ takes over the role of the branch $\alpha$.
\movieref{a}
}
  \label{fig:res_3_1_1}
\end{figure}

Bifurcations with $|\mT| = 1$ cause prominent bends in a family of elliptic
\onetori{} on either side of the crossing resonance, which leads to pronounced
gaps. While the elliptic branch on one side of the gap just bends, the one on
the other side gets hyperbolic at some point. The point where this family
changes its stability is similar to a \twoD{} saddle-node bifurcation.
In the \fourD{} case this ``new'' elliptic family continues the family of
\onetori{} that stays elliptic
beyond the gap. The $|\mT| = 1$ bifurcations lead to bends in frequency
space, which has also been observed for \threeD{} volume-preserving
maps~\cite{DulMei2009}.

As an example the $3:1:1$ resonance intersecting \Mone{} is considered, because
it causes a prominent bend in phase space and in frequency space, marked by
\bboxed{1} in Figures~\ref{fig:phase_space} and
\ref{fig:freq_space}. Note that also all the
smaller bends of \Mone{} and \Mtwo{} visible in
\prettyref{fig:phase_space} are
caused by crossing resonances with $|\mT| = 1$.

Figure~\ref{fig:res_3_1_1_ps} shows the family
\Mone{} (orange) at the bend with the hyperbolic branch (green) in a \pss{}.
Coming from the top of \prettyref{fig:res_3_1_1_ps} the upper
elliptic branch
$\alpha$ (orange) stays elliptic and bends to the right.
Below that another elliptic branch $\beta$ and a hyperbolic branch emerge from
a saddle-node bifurcation. In particular, the elliptic branch $\beta$ continues
the elliptic branch $\alpha$ beyond the gap.
This observation is consistent with the normal form
results~\cite{BroHanJorVilWag2003} shown in Figs.~\ref{fig:res_3_1_1}(b)--(d).
The stable equilibrium in \prettyref{fig:res_3_1_1_normform1} corresponds to the
upper branch labeled with $\alpha$ in \prettyref{fig:res_3_1_1_ps}. The plot
(c) shows the normal form just after the saddle-node bifurcation which creates
the hyperbolic and the lower elliptic branch $\beta$. The last plot (d)
visualizes how the elliptic branch $\beta$ moves to the center and takes over
the role of the original elliptic family $\alpha$.

\subsection{$\boldmath{\mT = 0}$ \label{sec:mN0}}

The condition $\mT = 0$ means that the longitudinal frequency $\nuL$ of a
family of \onetori{} crosses a rational value $n/\mL$.
This leads to the break-up of the resonant elliptic \onetorus{} into a chain of
periodic orbits of alternating \EE{} and \EH{} stability. This looks like the
well-known \twoD{} Poincar\'e-Birkhoff scenario~\cite{LicLie1992} embedded in a
\twoD{} manifold composed of a family of elliptic \onetori{}.

\begin{figure}[tb]
  \includegraphics[width=7cm]{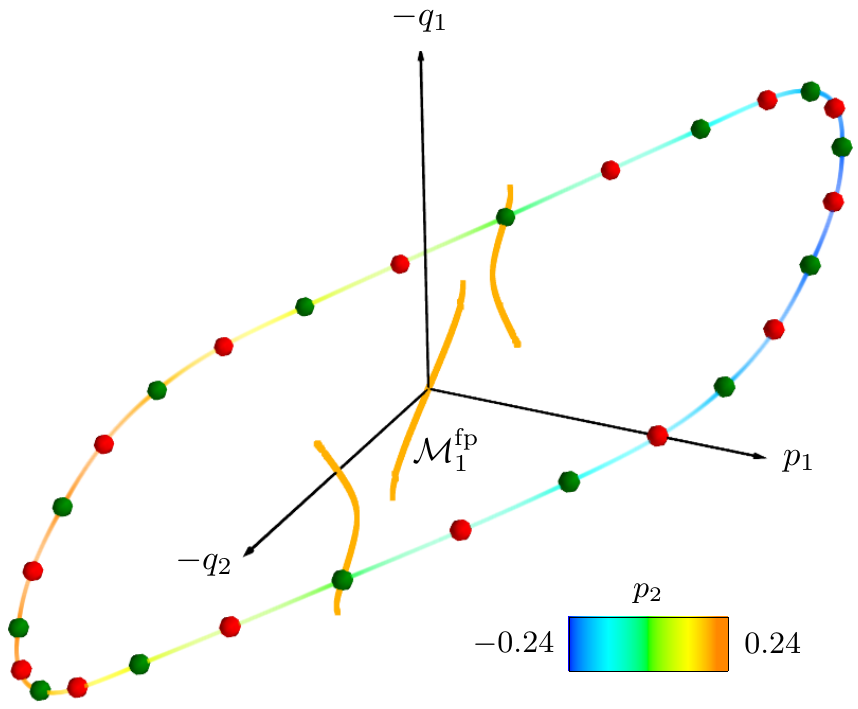}
  \caption{Chain of periodic orbits of type \EE{} (red) and \EH{} (green),
which are the remnants of a \onetorus{} of \Mone{} that fulfilled the
$7:\bboxed{0}:2$
resonance. In addition the points in the \pss{} of the family \Mone{} and a
\threeD{} projection of a close-by \onetorus{} with the fourth coordinate
encoded as color is shown for
comparison.
\movierefall{}
  }
  \label{fig:res_7_0_2_ps}
\end{figure}

We consider the $7:0:2$ resonance as example, marked by \bboxed{0}
in Figures~\ref{fig:phase_space} and \ref{fig:freq_space}.
This case was already
discussed in detail~\cite{LanRicOnkBaeKet2014} as limiting
case of an uncoupled rank-1-resonance. The resulting chain of \EE{} and
\EH{} periodic orbits and a close-by \onetorus{} are shown in a color
projection~\cite{PatZac1994} onto the \pss{} in
\prettyref{fig:res_7_0_2_ps}. For orientation, the orange
points of the main
family \Mone{} from \prettyref{fig:phase_space} are included.

Note that this is the only type of resonance which also affects families of
hyperbolic \onetori{} as they just have a longitudinal frequency \nuL{}.
Hence, this is the only way to get deeper into a hyperbolic hierarchy:
If the longitudinal frequency crosses a rational value, the resonant hyperbolic
\onetorus{} breaks up in a chain of alternating \HH{} and \EH{} periodic
points. The \EH{} periodic points have again a family of hyperbolic \onetori{}
attached~\cite{JorVil1997}.
Like in the elliptic case the dynamics restricted to the manifold composed of a
family of hyperbolic \onetori{} looks similar to a \twoD{} Poincar\'e-Birkhoff
scenario.

\section{Global behavior \label{sec:global}}

In the previous section the local structure
of bifurcations of families of \onetori{} has been considered.
Now we want to discuss the global behavior, i.e.\
the geometry and the properties of the newly
created \onetori{} far away from the bifurcation.
Furthermore, we discuss the geometry of a family of \onetori{}
whose corresponding resonance line intersects both main families $\Mone$ and
$\Mtwo$.

\subsection{Properties of bifurcated families of 1D-tori
\label{sec:global_props}}

\begin{figure}[tb]
\includegraphics[width=8.5cm]{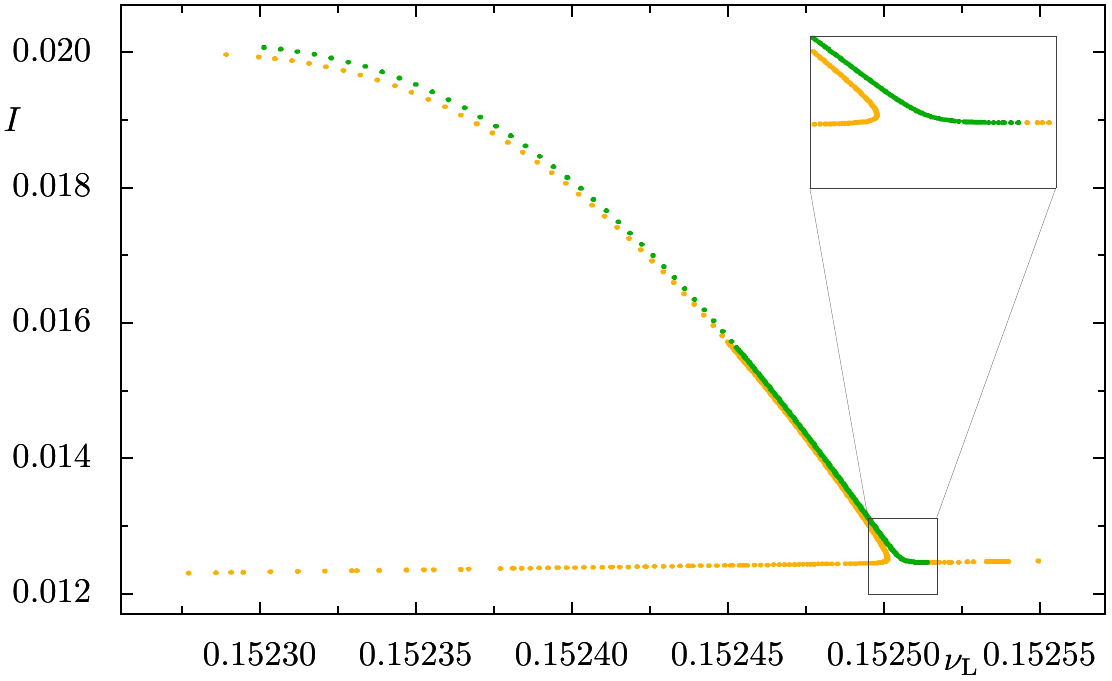}
  \caption{Action-frequency plot for the $-1:15:2$ resonance crossing the main
family \Mtwo{}. The elliptic \onetori{} are shown in orange and the hyperbolic
ones in green.}
  \label{fig:res_-1_15_2}
\end{figure}

Near the bifurcation, i.e.\ near the crossing of a resonance line and a main
family, the underlying geometry is described by the corresponding normal form.
On the other hand, further
away from the bifurcation, pairs of elliptic and hyperbolic \onetori{} can
be thought of as remnants of broken resonant \twotori{}, which has been
shown~\cite{Tod1994} using normal form analysis for resonances of order $\geq5$
and illustrated for a generic \fourD{} map~\cite{LanRicOnkBaeKet2014}. Note
that this is analogous to the well-known Poincar\'e-Birkhoff scenario in
\twoD{} maps.
The break-up of a resonant \twotorus{} suggests that the remnant elliptic and
hyperbolic \onetori{}
have similar longitudinal frequency \nuL{} and a similar action $I =
\oint_{\gamma}\sum_{i=1,2} p_i \mathrm{d}q_i$ with $\gamma$ being a
path along the remnant \onetorus{}.
Moreover, both \onetori{} resemble the geometry of the original \twotorus{},

As an illustration we consider the $-1:15:2$ resonance,
see \prettyref{fig:freq_space}, for which an action-frequency plot is shown in
\prettyref{fig:res_-1_15_2}. The bifurcation
with $|\mT| = 1$ takes place at $\nuL \approx 0.152506$ and a pair of an
elliptic and a hyperbolic family of \onetori{} branches off the main family
\Mtwo{} with increasing action. Their two actions $I(\nuL)$ match remarkably
well far away from the bifurcation. An even better agreement was
observed~\cite{Onk2015} for generic bifurcations, i.e.\ $|\mT| \geq 5$.

In addition, we also find a relation in the normal behavior. For bifurcations
with $|\mT| \geq 5$ we observe for a pair of \onetori{}
with equal values of \nuL{} that $2 \pi$ times
the normal frequency $\nuT$ of the elliptic \onetorus{} and the
Lyapunov-exponent $\lamN{}$ of the hyperbolic one match surprisingly well even
far away from the bifurcation. The relation of \nuT{} and \lamN{} close to
the bifurcation can be explained using the normal form expression for the
$|\mT| \geq 5$ bifurcation. From the linearized dynamics,
see e.g.\ Eq.~(5.4.19) in Ref.~\cite{LebMou1999},
it follows that $2 \pi \nuT{} = \lamN{}$.
In addition, we also observe for $|\mT| = 1$ bifurcations a very good match of
$\lamN{}$ and $2 \pi \nuT$ away from the bifurcation.

As discussed in the previous section, the families of \onetori{}
bifurcating from the main family when crossing a resonance are
the skeleton of the corresponding resonance channel. Our findings
suggest that one can estimate the properties of a hyperbolic
\onetorus{} within such a resonance channel from its elliptic
counterpart, which is numerically much easier to compute. This is especially
interesting as the Lyapunov-exponents are relevant for the chaotic transport
within the channel.

\subsection{Connection of bifurcations}

\begin{figure}[htbp]
  \subfloat[][]{
    \includegraphics{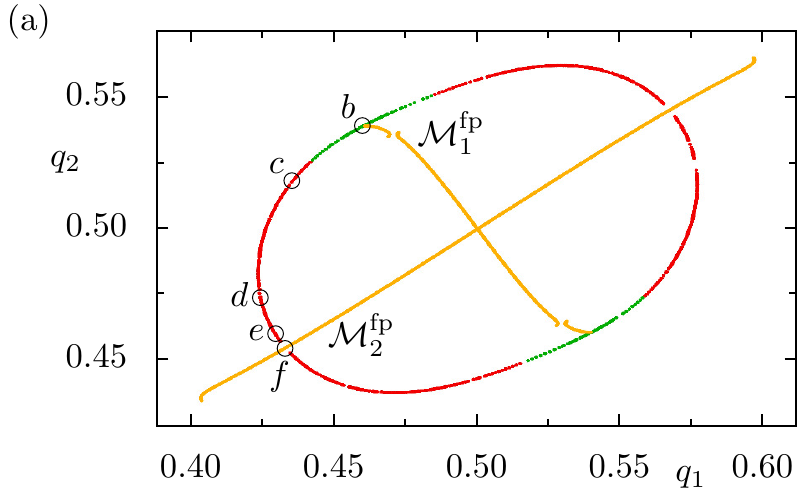}
    \label{fig:res_10_0_3_q1q2}
  } \\
  \subfloat[][]{
    \includegraphics{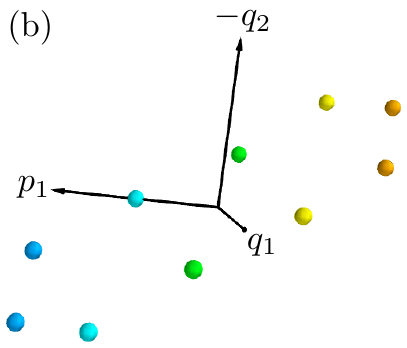}
  }
  \subfloat[][]{
    \includegraphics{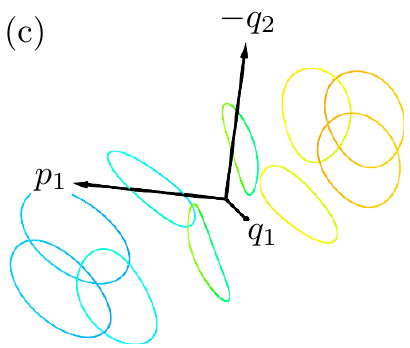}
  } \\
  \subfloat[][]{
    \includegraphics{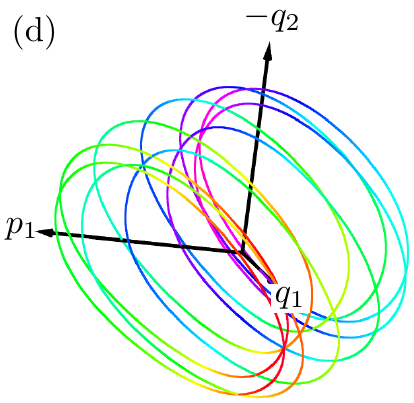}
  }
  \subfloat[][]{
    \includegraphics{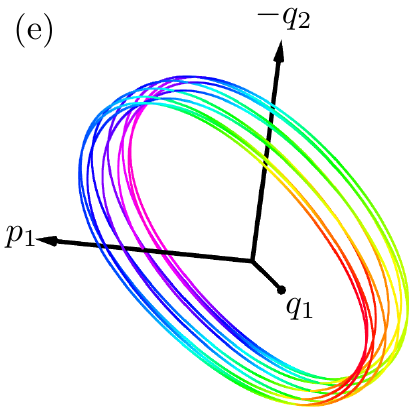}
  } \\
  \subfloat[][]{
    \includegraphics{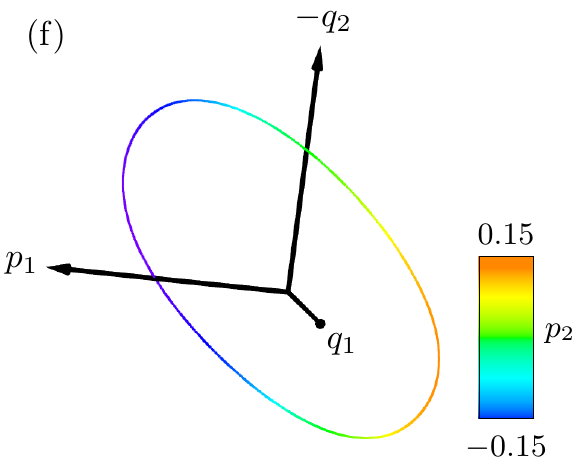}
  }
  \caption{Visualization of $10:0:3$ resonance connecting both main families.
(a) The main families \Mone{}, \Mtwo{} (orange) and one of the two connecting
families of \onetori{} (elliptic -- red, hyperbolic -- green) are shown. The
series (b)--(f) shows the \threeD{} projections of \onetori{} at the positions
marked in (a) with the $p_2$-coordinate encoded as color.
\movierefall{}}
  \label{fig:res_10_0_3}
\end{figure}

An interesting global connection happens if a resonance crosses both main
families \Mone{} and \Mtwo{}.
While the local behavior at each crossing point is described by the normal
form of the corresponding bifurcation, the geometry of such connections is
non-trivial. For example, consider the $m_1:m_2:n = 10:0:3$ resonance in
\prettyref{fig:freq_space} for which \mT{} is
given by $m_2 = 0$ at the intersection with \Mone{} while at the
intersection with \Mtwo{} it is given by $m_1 = 10$.
Hence, the coefficients $m_1, m_2$ exchange their roles along the
family of \onetori{}.
This implies, that the longitudinal and the normal radius, see
\prettyref{fig:torus}, of the original, resonant
\twotori{} interchange along the resonance, i.e.\ the normal radius increases
and the longitudinal radius decreases.

We now illustrate the geometry of this connection by a visualization in
phase space. For this we use the \pss{}, which in case of the considered
${10:0:3}$ resonance can be reduced further to a plot in the $(q_1-q_2)$-plane.
This is possible as the $p_1$ coordinate is zero for all points lying in the
\pss{} due to symmetry, see \prettyref{app:symm_breaking}.
Figure~\ref{fig:res_10_0_3_q1q2} shows one of the two families of \onetori{}
(green and red points) of the $10:0:3$ resonance connecting the main families
\Mone{} and \Mtwo{} (orange).

The change in geometry of the \onetori{}
when going from \Mone{} to \Mtwo{} along the resonance is
illustrated in \prettyref{fig:res_10_0_3}(b)--(f) by color
projections,
i.e.\ the fourth coordinate $p_2$ is encoded in color.
When the resonance crosses \Mone{} it causes a $|\mT| = 0$ bifurcation,
i.e.\ the \onetorus{} with $\nuL = 3/10$ breaks up in a chain of
\EE{} and \EH{} periodic points as described in
\prettyref{sec:mN0}, see \prettyref{fig:res_10_0_3}(b). At the
\EE{} periodic
points two elliptic families of \onetori{}
and at the \EH{} periodic points one hyperbolic family of \onetori{} are
attached.
One of the two elliptic families is embedded in \Mone{}, see Fig.~4c in
Ref.~\cite{LanRicOnkBaeKet2014}, while
the other elliptic family of \onetori{} is, as in \prettyref{sec:global_props},
the counterpart to the hyperbolic family of \onetori{}.
Following the hyperbolic family away from the bifurcation the geometry of the
broken
\twotorus{} becomes visible. For instance, the \onetorus{} shown in
\prettyref{fig:res_10_0_3}(c) consists of ten disjoint rings,
which are
arranged like an orbit on a \twotorus{} with resonant longitudinal frequency
$\nuL = 3/10$.

With increasing distance from the family \Mone{} the rings get larger and start
to overlap in the \threeD{} projection, see
\prettyref{fig:res_10_0_3}(d).
This continues until the ten rings are
oriented in the longitudinal direction like an orbit on a \twotorus{} with
resonant normal frequency $\nuT = 3/10$, see
\prettyref{fig:res_10_0_3}(e).
Finally, all rings collapse onto one \onetorus{} of \Mtwo{}, see
\prettyref{fig:res_10_0_3}(f).
This corresponds to a type $|\mT| = 10$ bifurcation caused by the $10:0:3$
resonance crossing \Mtwo{}.
Thus, the connection between \Mone{} and \Mtwo{} along the resonance
can be seen as exchange of the longitudinal and the normal radius of the
original \twotori{}.

Note that we observe a surprising change of stability along the connection
shown in \prettyref{fig:res_10_0_3_q1q2}. The connecting family starts
hyperbolic (green) at the \EH{} periodic orbit (b) and later changes stability
to elliptic (red). At the same point the corresponding elliptic family
of \onetori{} turns hyperbolic.
Thus one could call this a transcritical bifurcation of a \onetorus{}
because an exchange of stability occurs along the family.
Whether this is a generic phenomenon for such connections or specific to the
investigated resonance and system requires further investigation.

\section{Summary and outlook \label{sec:summary}}

In this paper we investigate the bifurcations of families of elliptic
\onetori{} which occur when they cross resonances.
We observe all stages of the bifurcations in the phase space of one \fourD{} map
without varying any parameter. This is possible because the
longitudinal and normal frequencies of the
families vary smoothly along a family.
As regular \twotori{} are arranged around
the skeleton of families of elliptic \onetori{}
their bifurcations leading to gaps, bends, and new branches
crucially determine the organization of the regular structures
in phase space.
To visualize the structures we use the frequency space, \psss{}, and local
\twoD{} projections of the \psss{}.

On a resonance the longitudinal and the normal frequency fulfill a
condition $\mL \cdot \nuL + \mT \cdot \nuT = n$. The crucial
number determining the type of the bifurcation is \mT{}.
If a family of elliptic \onetori{} crosses a resonance with $|\mT| = 1$,
it will form prominent bends.
A $|\mT| = 2$ bifurcation leads to an intermediate change of
stability and hence to a gap in the regular structures. At each end of the gap
a new family emerges.
The most complicated structure is caused by $|\mT| = 3$
bifurcations. Before the bifurcation a pair
of families of elliptic and hyperbolic \onetori{} is created
from which the hyperbolic family collapses onto the main family and re-emerges
afterwards with opposite orientation.
For generic bifurcations with
$|\mT| \ge 4$ one elliptic and one hyperbolic family of \onetori{} emerges
directly from the main family.
In the \pss{} the new families appear as $2 \cdot \mT{}$ branches of
alternating stability.
Using local \twoD{} projections we demonstrate that all the results are
consistent with normal form predictions and numerically support the conjecture
that they are also valid for generic \fourD{} maps~\cite{BroHanJorVilWag2003}.
In terms of these \twoD{} projections the bifurcations in \fourD{} maps
resemble bifurcations of fixed points in \twoD{} maps.

The families of \onetori{} emerging from a bifurcation form the
skeleton of the corresponding resonance channel.
We investigate the families along a resonance channel even far away from the
bifurcation.
For bifurcations with $|\mT| \ge 4$ we find an interesting relation between the
new elliptic and hyperbolic families:
for each pair of elliptic and hyperbolic \onetori{} the normal frequency \nuT{}
and the Lyapunov exponent \lamN{} match surprisingly well,
$2 \pi \nuT \approx \lamN$.
Thus, based on the elliptic family it is possible to gain
information on the hyperbolic family, which is numerically
much harder to find and to compute.
Moreover, for resonances intersecting the two main families
of \onetori{} an interesting, non-trivial geometric
connection is observed.

There are also several interesting
open questions and future applications: Where and why do families of
elliptic and hyperbolic \onetori{} end?
How do bifurcations of periodic points connect
with the picture obtained from the bifurcations of families of
\onetori{}?
Furthermore, based on the skeleton of a resonance channel one can
investigate how the stable and unstable manifolds
of the families of hyperbolic \onetori{} govern the chaotic transport.
Moreover, understanding the geometry of junctions of resonance channels is of
importance as they play an important role for chaotic transport in the Arnold
web.

\section*{Acknowledgements}

We are grateful for discussions with Henk Broer, Holger Dullin, {\`A}lex Haro,
\`{A}ngel Jorba, Srihari Keshavamurthy, Alejandro Luque, and
Jim Meiss.
Furthermore, we acknowledge support by the Deutsche Forschungsgemeinschaft
under grant KE~537/6--1.

All \threeD{} visualizations were created using
\textsc{Mayavi}~\cite{RamVar2011}.

\appendix

\section{Symmetry breaking bifurcations \label{app:symm_breaking}}

For bifurcations of periodic orbits in \twoD{} area-preserving maps
it is well-known that the presence of symmetries leads to additional types
of bifurcations~\cite{Rim1978, AguMalBarDav1987, MaoDel1992}.
In this section we demonstrate that symmetries also cause new types of
bifurcations of families of \onetori{} in addition to the ones described in
\prettyref{sec:bifs}. Moreover, we discuss further implications of the
presence of symmetries for families of \onetori{}.

The symmetries of a variant of the reversible~\cite{RobQui1992} \fourD{} map \prettyref{eq:map},
were already studied~\cite{KooMei1989a}.
From this one obtains the reversing symmetry operator
\begin{equation}
\label{eq:S1}
  \begin{aligned}
    S_1: && p' &= p &\\
            && q' &= -q - p&
  \end{aligned}
\end{equation}
with $p = (p_1, p_2)$, $q = (q_1, q_2)$, for the map \prettyref{eq:map}.
A reversing symmetry~\cite{Lam1992} is an involution, $S_1^2 = \text{Id}$, and allows for a simple
expression of the inverse map, $f^{-1} = S_1 \circ f \circ S_1$.
Using an inversion, i.e.\ changing the sign of all coordinates, one obtains a
further reversing symmetry~\cite{PinJim1987}
\begin{equation}
\label{eq:S2}
  \begin{aligned}
    S_2: && p' &= - p & \\
        && q' &= q + p &.
  \end{aligned}
\end{equation}
As an example we explain the unusual bifurcations corresponding to the
resonances $-1:2:0$ and $-1:10:1$ shown in \prettyref{fig:freq_space} by these
two symmetries.

For the $-1:2:0$ resonance intersecting \Mtwo{} with $|\mT{}| = 1$
one would expect, according to \prettyref{sec:mN1},
a strong bend of the family in phase space.
Instead, in phase space a large gap with new branches at each end and in
frequency space prominent bends are observed.
Thus, this corresponds to the geometry found for the case
$|\mT{}| = 2$, see \prettyref{sec:mN2}.
However, for the $-1:2:0$ resonance the new branches are not dynamically
connected, i.e.\ twice the number of families arise.
The symmetry operator $S_1$ provides the mapping between
these two families. This demonstrates the symmetry-breaking
nature of this bifurcation, similar to the symmetry-breaking period-doubling of
periodic orbits in area-preserving maps.
Note that for $|\mT{}| = 1$ we only observe symmetry breaking when
$\mL{}$ is even.

For the $-1:10:1$ resonance intersecting \Mtwo{}
one would also expect the $|\mT{}| = 1$ behavior. Instead, the geometry in
phase space even looks like a $|\mT{}| = 4$ bifurcation,
i.e.\ in a \pss{} four elliptic branches emerge directly from the main family.
Again these branches are not dynamically connected:
one pair of them is related by $S_1$, while the other pair is
related by $S_2$.
Hence, this is a double symmetry breaking bifurcation of a family of \onetori{}.
Note that the corresponding four families of hyperbolic \onetori{} were not
found due to technical difficulties.

The presence of symmetries has further implications for families of \onetori{}:
Consider a \onetorus{} with longitudinal frequency $\omega$, which can be
represented by $x(\theta)$ with $f^{\pm 1} (x(\theta)) = x(\theta\pm\omega)$ and
$\theta \in [0,2\pi)$, see \prettyref{eq:map=rot}. Such a \onetorus{} is mapped
by a symmetry $S$ to a \onetorus{} with same frequency, as from $f^{\pm 1} = S
\circ f^{\mp 1} \circ S$ and $S^2 = \text{Id}$ follows $(f^{\mp 1} \circ
S)(x(\theta)) =  S(x(\theta\pm\omega))$. For instance, each of the families
\Mone{} and \Mtwo{} is mapped by a symmetry to a family of \onetori{}. As the
families \Mone{} and \Mtwo{} are the only families which contain the \EE{} fixed
point \fixedpoint{}, which is invariant under the symmetries $S_1$ and $S_2$,
there are only two options: either $S_i(\Mone{})=\Mone{}$ and
$S_i(\Mtwo{})=\Mtwo{}$, or $S_i(\Mone{})=\Mtwo{}$ and $S_i(\Mtwo{})=\Mone{}$.
As the range of frequencies $\omega$ of these families \Mone{} and
\Mtwo{} does not coincide, see \prettyref{fig:freq_space}, the first option
holds, i.e.\ the families \Mone{} and \Mtwo{} are invariant under $S_1$ and
$S_2$. Furthermore, we observe that every \onetorus{} of these families has
exactly two intersection points $x_1, x_2$ with the \threeD{} slice $p_2 = 0$.
The points $S_1(x_1)$, $S_1(x_2)$, $S_2(x_1)$, $S_2(x_2)$ also have $p_2=0$,
since $|p|$ is conserved under $S_1$ and $S_2$, see Eqs.~(\ref{eq:S1}) and
(\ref{eq:S2}). Thus, this set of points has to coincide with the set of
points $\{x_1, x_2\}$. This is only possible if the
intersection points are of the form $x_1=(0,0,q_1,q_2)$, $x_2=(0,0,-q_1,-q_2)$.
For this reason the families \Mone{} and \Mtwo{} intersect
with the \twoD{} plane defined by $p_1 = p_2 = 0$ on a line rather than just in
a point. To show this for other families
of \onetori{} similar arguments may be applied.

\section{Computation of 1D-tori \label{app:algo}}

To compute invariant \onetori{} we use a Fourier expansion
method~\cite{CasJor2000, JorOll2004}. In the following we will give a
brief review~\cite{Onk2015} to comment on specific choices of parameters used
in this paper.
The basic idea of the method is to describe a \onetorus{} by a finite Fourier
series
\begin{equation}
x(\theta) = a_0 + \sum_{k=1}^N a_k \cos(k\theta) + \sum_{k=1}^N b_k
\sin(k\theta)
\label{eq:x_theta}
\end{equation}
with $4$-dimensional coefficients $a_k$, $b_k$ and an angle $\theta
\in [0, 2\pi)$. For highly deformed \onetori{} we used up to $N =
240$ coefficients.

On every invariant 1D curve with longitudinal frequency $\omega = 2
\pi \nuL{}$ one has
\begin{equation}
f(x(\theta)) = x(\theta + \omega),
\label{eq:map=rot}
\end{equation}
where $f$ denotes the map~\prettyref{eq:map}.
Thus one can define an error of the approximation by
\begin{equation}
F_N (x(\theta_i)) = f(x(\theta_i)) - x(\theta_i + \omega)
\label{eq:def_F}
\end{equation}
which is evaluated on the grid $\theta_i = 2 \pi i/(2N+1)$, $i \in
\{0, 1, \dots, 2N\}$.
For a fixed $\omega$ a high-dimensional Newton search in the coefficients
$a_k$ and $b_k$ is used to find zeros
of $F_N(x(\theta))$. We stop the search if an error of $||F_N(x(\hat
\theta))||_{\infty} < 10^{-12}$ is reached where $\hat \theta$ denotes
a 10 times finer grid than $\theta$.
Note that the invariant curve described by $x(\theta)$ is the same object as
$x(\theta + \varphi)$ for any $\varphi \in \mathbb{R}$. To get rid of this
freedom, one needs to add a uniqueness
condition. As we want to visualize the \onetori{} in a \pss{} with
$p_2^* = 0$, we use $p_2 = 0$ as this condition.

For the Newton search an initial guess for the coefficients $a_k$, $b_k$ and the
longitudinal frequency $\omega$ is needed. If at least one \onetorus{} of
a family of \onetori{} is known, an initial guess for a close-by \onetorus{} of this
family can be obtained by extrapolating within
the ($4(2N +1)+1$)-dimensional space spanned by $a_k$, $b_k$ and $\omega$.
As continuation parameter the Euclidian distance in this space is used~\cite{CasJor2000}.
Then, the whole family can be computed using a predictor-corrector
continuation. Thus, the crucial task is to either find an initial \onetorus{}
of the family and compute its Fourier coefficients or to obtain a guess for the
Fourier coefficients from geometric considerations. This is described in the
following by three approaches:

(i) The first approach is to start from \twotori{} and apply the
geometric contraction method~\cite{LanRicOnkBaeKet2014}
to obtain elliptic \onetori{}.
This is particularly useful as it allows for obtaining
\onetori{} beyond bends and large gaps where a direct continuation
method usually fails.

(ii) For families attached at \EE{} or \EH{} periodic points $\vec{u}$ one can use
each pair of elliptic eigenvalues $\exp{(\pm \ui \ 2\pi \nu)}$ and the
corresponding eigenvectors $\vec{\xi}, \bar{\vec{\xi}}$ of the linearized map at
the periodic point to define an initial guess, i.e.\ $a_0 = \vec{u}$, $a_1 = 2
\varepsilon \Re (\vec{\xi} \,)$, $b_1 = 2\varepsilon \Im(\vec{\xi} \,)$,
$\omega = 2 \pi \nu + \Delta \omega$ with small $|\varepsilon|$ and $|\Delta
\omega|$.

(iii) The previous two approaches do not work for families of hyperbolic \onetori{}
arising from bifurcations. For bifurcations with odd $|\mT{}|$ these
families are computed by extrapolating the elliptic branches through the main
family, see \prettyref{fig:res_5_-5_1_ps}. For even $|\mT{}|$
we use the following approach~\cite{Onk2015}:
From \prettyref{sec:global_props} it is known that the longitudinal frequencies
of the new elliptic and hyperbolic families
match. Thus, we can choose a rational value $n/m$
in their frequency range and find the corresponding \EH{} or \HH{} $m$-periodic
orbit which is embedded in the hyperbolic family according to
\prettyref{sec:mN0}.
As these periodic orbits roughly lie on a \onetorus{}, we approximate
a nearby \onetorus{} by an ellipse using the mean of the periodic points
$\vec{C}$ and two of the periodic points $\vec{A}$ and $\vec{B}$, i.e.\ $a_0 =
\vec{C}$, $a_1 = \vec{A} - \vec{C} + \vec{\delta}_a$, $b_1 = \vec{B} - \vec{C} +
\vec{\delta}_b$, $\omega = n/m + \Delta \omega$ with small
$||\vec{\delta}_b||$, $||\vec{\delta}_a||$, and $|\Delta \omega|$.

\end{document}